\documentclass[11pt]{article}
\usepackage{amsmath,amssymb,amsfonts,epsf,graphicx}
\usepackage[nosort]{cite}
\usepackage[margin=1in]{geometry}

\makeatletter \@addtoreset{equation}{section} \makeatother

\newcommand{\fft}[2]{{\frac{#1}{#2}}}
\newcommand{\ft}[2]{{\textstyle\frac{#1}{#2}}}

\let\C=\Chi

\def\nn{\nonumber}

\let\bm=\bibitem

\newcommand{\be}{\begin{equation}}
\newcommand{\ee}{\end{equation}}
\def\ba{\begin{array}}
\def\ea{\end{array}}
\def\ft#1#2{{\textstyle{\frac{\scriptstyle #1}{\scriptstyle #2}}}}
\def\fft#1#2{\frac{#1}{#2}}

\def\sst#1{{\scriptscriptstyle #1}}

\def\td{\tilde}

\def\dalemb#1#2{{\vbox{\hrule height .#2pt
        \hbox{\vrule width.#2pt height#1pt \kern#1pt
                \vrule width.#2pt}
        \hrule height.#2pt}}}
\def\square{\mathord{\dalemb{6.8}{7}\hbox{\hskip1pt}}}

\newcommand{\bea}{\begin{eqnarray}}
\newcommand{\eea}{\end{eqnarray}}

\def\0{{\sst{(0)}}}
\def\1{{\sst{(1)}}}
\def\2{{\sst{(2)}}}
\def\3{{\sst{(3)}}}
\def\4{{\sst{(4)}}}
\def\5{{\sst{(5)}}}
\def\6{{\sst{(6)}}}
\def\7{{\sst{(7)}}}
\def\8{{\sst{(8)}}}

\def\ep{{\epsilon}}

\def\R{\rlap{\rm I}\mkern3mu{\rm R}}

\def\R{{{\mathbb R}}}
\def\C{{{\mathbb C}}}

\def\CP{{{\mathbb C}{\mathbb P}}}

\def\Z{{{\mathbb Z}}}

\begin{document}

\begin{flushright}
\hfill{ \
NSF-KITP-12-145\ \ \ \ }
\end{flushright}

\begin{center}\ \\ \vspace{50pt}
{\Large {\bf Brane Resolution Through Fibration}}\\ 
\vspace{30pt}

Justin F. V\'azquez-Poritz$^{1,2,3}$ and Zhibai Zhang$^{1,2}$
\vspace{20pt}

$^1${\it Physics Department\\ New York City College of Technology, The City University of New York\\ 300 Jay Street, Brooklyn NY 11201, USA}

\vspace{10pt}
$^2${\it The Graduate School and University Center, The City University of New York\\ 365 Fifth Avenue, New York NY 10016, USA}

\vspace{10pt}
$^3${\it Kavli Institute for Theoretical Physics\\ University of California, Santa Barbara CA 93106, USA}\\

\vspace{20pt}

{\tt jvazquez-poritz@citytech.cuny.edu; zzhang2@gc.cuny.edu}

\end{center}

\vspace{30pt}

\centerline{\bf Abstract}

\noindent We consider $p$-branes with one or more circular directions fibered over the transverse space. The fibration, in conjunction with the transverse space having a blown-up cycle, enables these $p$-brane solutions to be completely regular. Some such circularly-wrapped D3-brane solutions describe flows from $SU(N)^3$ ${\cal N}=2$ theory, ${\mathbb F}_0$ theory, as well as an infinite family of superconformal quiver gauge theories, down to three-dimensional field theories. We discuss the operators that are turned on away from the UV fixed points. Similarly, there are wrapped M2-brane solutions which describe smooth flows from known three-dimensional supersymmetric Chern-Simons matter theories, such as ABJM theory. We also consider $p$-brane solutions on gravitational instantons, and discuss various ways in which U-duality can be applied to yield other non-singular solutions.

\thispagestyle{empty}

\pagebreak

\tableofcontents
\addtocontents{toc}{\protect\setcounter{tocdepth}{3}}
\newpage


\section{Introduction and summary}

Supergravity $p$-brane solutions play an important role in string theory, and have provided many explicit examples of the AdS/CFT correspondence \cite{agmoo}. A singularity at the origin of the brane would impose a severe restriction on the range of validity of the supergravity background. While some singularities may be resolved by stringy or non-perturbative effects, deformations of the geometry and/or the inclusion of additional fields can sometimes smooth out the singularity at the level of supergravity. 

For instance, this can be done through Chern-Simons type modifications in the Bianchi identities or equations of motion. There is an abundance of literature on this type of singularity resolution of $p$-brane solutions, some examples of which can be found in \cite{KS,transgression,pope1,cvetic1}. This can provide supergravity backgrounds dual to supersymmetric field theories for which the conformal symmetry is typically broken by the resolution. The resolution procedure involves turning on an $n$-form which is square integrable at short distance, which requires the transverse space to have a non-collapsing $n$-cycle. As an example of this, consider the Klebanov-Strassler solution, which provides a supergravity description of chiral symmetry breaking and confinement \cite{KS}. This involves D3-branes on a deformed conifold, which has a non-collapsing 3-cycle and therefore supports a square-integrable three-form flux. 

We will consider the scenario in which the the $d$-dimensional transverse space has a non-collapsing 2-cycle and/or $d-2$-cycle. We wrap the $p$-brane on one or more circular directions which are fibered over the transverse direction. The fibration manifests itself as a harmonic 2-form source term for the form field that supports the $p$-brane. The non-collapsing cycle translates into there being a 2-form that is square integrable at short distance, which ends up yielding a non-singular $p$-brane solution. Previous examples of this construction include D5-branes on the Eguchi-Hanson and Taub-NUT instantons \cite{overlapping}, D3-branes on a resolved cone over $T^{1,1}/\Z_2$ \cite{lu1}, M2-branes on a regular cone over $Q^{1,1,1}/\Z_2$ \cite{chen}, and various branes on Taub-NUT and Taub-BOLT instantons \cite{Nuts-bolts}. 

Here we generalize these already-known solutions by considering multiple fibrations, as well as greatly enlarge the number of examples of toroidally-wrapped $p$-branes by considering various spaces with non-collapsing $d-2$-cycles. The new solutions include D3-branes on cones over $S^5/\Z_3$ and the $Y^{p,q}$ spaces; M2-branes on cones over $S^7/\Z_4$, the Stiefel manifold, $M^{3,2}$, and Sasaki-Einstein manifolds in general; various $p$-branes on the Schwarzschild instanton and its higher-dimensional analog. 

The large distance limit of some of the wrapped D3-brane solutions describe known four-dimensional superconformal field theories, including $SU(N)^3$ ${\cal N}=2$ theory \cite{kachru,agmoo}, ${\mathbb F}_0$ theory \cite{morrison}, and an infinite family of superconformal quiver gauge theories \cite{Benvenuti}. The linearized form of the solutions enables us to analyze the transformation properties and dimensions of the operators turned on as one flows away from the UV fixed point, and in certain cases some of these operators can be identified. In general, the fibration corresponds to turning on a dimension-two vector operator, a blown-up 2-cycle corresponds to a dimension-two scalar operator, and a blown-up 4-cycle corresponds to a dimension-six non-mesonic scalar operator \cite{benvenuti}. The full supergravity solutions describe smooth flows in which the theory undergoes compactification to three dimensions. Likewise, some of the wrapped M2-brane solutions describe smooth flows from known three-dimensional supersymmetric Chern-Simons matter theories, such as ABJM theory \cite{Aharony}.

We also discuss how the above wrapped $p$-brane solutions are related to other types of supergravity solutions. For instance, magnetically-charged AdS solitons in gauged supergravity \cite{Ross} can be dimensionally oxidized to certain wrapped $p$-brane solutions. We demonstrate this for the case of wrapped D3-branes on cones over the $Y^{p,q}$ spaces by using the consistent Kaluza-Klein reduction ansatz presented in \cite{Buchel}. Moreover, the $p$-brane solutions have several directions along which one can perform dimensional reduction or T-duality. For example, T-dualizing or reducing along a worldvolume direction that is fibered over the transverse space yields a $p$-brane solution that is modified by additional flux. At the same time, the fibration causes a direction in the transverse space to have a radius that is nowhere vanishing and which stabilizes at large distance. This ensures that performing T-duality or dimensional reduction along this transverse direction yields a non-singular solution as well.

We would like to note that an alternative way to construct completely non-singular supergravity solutions is to localize the $p$-brane at a point on the blown-up cycle of the transverse space, instead of a uniform distribution of $p$-branes as above. For the case of D3-branes on the resolved conifold, this results in a solution that smoothly interpolates from AdS$_5\times T^{1,1}$ at large distance to AdS$_5\times S^5$ at short distance. This describes an RG flow from the $SU(N)\times SU(N)$ ${\cal N}=1$ superconformal fixed point in the UV to the $SU(N)$ ${\cal N}=4$ superconformal fixed point in the IR, which has been confirmed from the behavior of the superpotential \cite{Murugan}. This construction has been generalized to cover all of the resolved cones over the $L^{a,b,c}$ spaces, yielding RG flows from AdS$_5\times L^{a,b,c}$ to AdS$_5\times S^5/\Z_k$ \cite{Cvetic-JFVP}.

This paper is organized as follows. In section 2, we consider D3-branes with worldvolume directions that are wrapped and fibered over six-dimensional spaces, including cones over $S^5/\Z_3$, $T^{1,1}/\Z_2$, $Y^{p,q}$ and various gravitational instantons. We emphasize the cases which describe smooth flows from known four-dimensional superconformal field theories to three-dimensional theories, and discuss the operators that are turned on away from the UV fixed point. In section 3, we consider toroidally-wrapped M2-branes on cones over various Sasaki-Einstein spaces, including regular cones over $S^7/\Z_4$, $Q^{1,1,1}/\Z_2$, the Stiefel manifold and $M^{3,2}$. Some of these solutions describe flows from known three-dimensional supersymmetric Chern-Simons matter theories. In section 4, we consider more examples of toroidally-wrapped brane solutions on gravitational instantons.

\section{Toroidally-wrapped D3-branes}

The D3-brane of type IIB supergravity is supported by the self-dual 5-form field strength, with a six-dimensional Ricci-flat transverse space. If the transverse space has a 2-cycle or 4-cycle, then we can construct a toroidally-wrapped D3-brane with one or more of the worldvolume coordinates fibered over the transverse space. In addition, a D3-brane with a fibered timelike direction can be interpreted as a rotating D3-brane. The
solution with maximum fibrations generally has the form \cite{lu1}
\bea\label{wrapd3gen} 
ds_{10}^2 &=& H^{-1/2} \eta_{\mu\nu} d\td x^{\mu} d\td x^{\nu} + H^{1/2}\, ds_6^2,\\
F_\5&=& d^4 \td x \wedge dH^{-1}-{\ast_6 dH}+({\ast_6 dA_\1^{\mu}})\wedge d\td x_{\mu} +
\fft16 \epsilon_{\mu\nu\rho\sigma}d\td x^{\mu}\wedge d\td x^{\nu}\wedge d\td x^{\rho}\wedge dA_\1^{\sigma},\nn
\eea
where $d\td x^{\mu}\equiv dx^{\mu}+A_\1^{\mu}$, $dA_\1^{\mu}=m^{\mu}L_\2^{\mu}$ ($\mu$ not summed) are harmonic 2-forms in the transverse space of the metric $ds_6^2$, $\ast_6$ is the Hodge dual with respect to
$ds_6^2$ and $\mu,\nu=0,\dots 3$. The equations of motion are satisfied, provided that
\be \label{lapmod} 
\square H = -\fft12 \sum_{\mu=0}^3 \left( m^{\mu} L_\2^{\mu}\right)^2,
\ee
where $\square$ is the Laplacian on $ds_6^2$. Note that the periods of the wrapped directions are governed by the connections on the fibers.

\subsection{On a regular cone over $S^5/\Z_3$}

A regular cone over the lens space $S^5/\Z_3$ is given by \cite{calabi}
\bea\label{cone1} 
ds_6^2 &=& f^{-1} dr^2+\fft{r^2}{9} f \left( d\psi-\ft32 \sin^2\theta \sigma_3\right)^2
+r^2 ds_{\CP^2}^2,\nn\\
f &=&1-\fft{b^6}{r^6},
\eea
where the metric on $\CP^2$ is given by
\be
ds_{\CP^2}^2=d\theta^2
+ \fft14 \sin^2\theta \left( \sigma_1^2+\sigma_2^2+\cos^2\theta \sigma_3^2\right),
\ee
and $\sigma_i$ are left-invariant 1-forms of $SU(2)$. If $\psi$ has a period of $2\pi$ and $r\ge b$, then this space is completely regular and has a short-distance geometry of $\R^2\times \CP^2$. The parameter $b$ is related to the volume of the 4-cycle $\CP^2$. Note that $S^5/\Z_3$ is a smooth space, which corresponds to the fact that we are working in a limit in which all twisted states in the string theory are heavy. The metric (\ref{cone1}) follows the general construction for Ricci-flat K\"ahler metrics in \cite{page}.

A 2-form that preserves the isometry of (\ref{cone1}) has the form
\be\label{L1}
L_\2 = u_0\ e^0\wedge e^5 + u_1\ e^1\wedge e^2+u_2\ e^3\wedge e^4, 
\ee
expressed in the vielbein basis
\bea e^0 &=& \fft{dr}{\sqrt{f}},\qquad e^1=\fft{r}{2} \sin\theta \sigma_2,\qquad e^2 = \fft{r}{2} \sin\theta \sigma_1,\qquad e^3 = r d\theta,\nn\\ 
e^4 &=& \fft{r}{2} \cos\theta \sin\theta \sigma_3,\qquad e^5=\fft{r}{3} \sqrt{f} \left( d\psi-\ft32 \sin^2\theta \sigma_3\right),
\eea
where the $u_i$ are functions of $r$ only. The K\"ahler form corresponds to $u_0=-1$, $u_1=u_2=1$. Another harmonic 2-form has
\be\label{form2}
u_0=\fft{2}{r^6},\qquad u_1=u_2=\fft{1}{r^6},
\ee
which corresponds to
\be
A_{(1)}=-\fft{m}{6r^4} \left( d\psi-\ft32 \sin^2\theta \sigma_3\right).
\ee
One of the worldvolume directions, let us suppose the $x_3$ direction, can be fibered over the transverse space. This yields a regular solution with
\be 
H=\fft{m^2}{4b^6 r^4}.
\ee
The connection on the fiber implies that $x_3$ has a period of $\pi m/(3b^4)$. The ten-dimensional metric interpolates between a product space of three-dimensional Minkowski spacetime and a $U(1)$ bundle over $\R^2\times \CP^2$ at short distance to AdS$_5\times S^5/\Z_3$ at large distance. 

The UV limit of the dual field theory is an ${\cal N}=2$ superconformal field theory of a general class of theories \cite{kachru} which was discussed specifically in \cite{agmoo}. It is an $SU(N)^3$ gauge theory with chiral multiplets $U_j$ in the $(\bf N,\bf\bar N,\bf 1)$ representation, $V_j$ in the $(\bf 1,\bf N,\bf\bar N)$ representation and $W_j$ in the $(\bf\bar N,\bf 1,\bf N)$ representation, where $j=1,2,3$. The classical superpotential has the form $W=g\epsilon^{ijk} U_i V_j W_k$, for which all three gauge couplings and the superpotential coupling $g$ are equal. In the quantum theory, the space of these couplings contains a one-dimensional line of superconformal fixed points whose parameter can be identified with the dilaton in the AdS$_5\times S^5/\Z_3$ background.

While mesonic directions of the full moduli space correspond to the motion of the D3-branes, the non-mesonic, or baryonic, directions are associated with either deformations of the geometry or turning on $B$-fields \cite{benvenuti}. As is generally the case for deformations which correspond to blowing up 4-cycles, $b$ is associated with a local deformation, since it does not change the position of the branes at infinity. Thus, as one flows away from a superconformal fixed point of the theory, non-mesonic operators get vacuum expectation values. The transformation properties and dimensions of these operators can be read off from the linearized form of the supergravity solution. In particular, it can be seen from (\ref{cone1}) that the leading order contribution of the $b$ parameter to the metric goes as $b^6/r^6$. This implies that a dimension-six non-mesonic scalar operator gets a vacuum expectation value that goes as $b^6$.

Although the specific operator has not been identified, it has been suggested that for this type of deformation the operator being turned on is associated with the gauge groups in the quiver and has the schematic form \cite{benvenuti}
\be\label{op1}
{\cal O}_i=\sum_g c_{i,g} {\cal W}_g \bar {\cal W}_g,
\ee
where the gauge groups have been summed over, ${\cal W}_g$ is an operator associated with the field strength for the gauge group $g$, and $c_{i,g}$ are constants. In addition, this operator could have a contribution from the chiral fields of the form
\be\label{op2}
U_i \bar U^i V_j \bar V^j W_k \bar W^k.
\ee
It is proposed that a particular combination of the terms in (\ref{op1}) and (\ref{op2}) corresponds to the blown-up 4-cycle.

In order to analyze the operator that corresponds to the fibration, it is useful to express the metric as
\be
ds_{10}^2 = \fft{r^2}{R^2} \left[ -dt^2+dx_1^2+dx_2^2+f dx_3^2\right]+\fft{R^2}{r^2 f} dr^2
+ R^2 \left[
\fft{1}{9} \left( d\psi-\ft32 \sin^2\theta \sigma_3-\fft{6b^6}{mr^2} dx_3\right)^2 + ds_{\CP^2}^2\right],\nn
\ee
where $R^2=m/(2b^3)$. Now one can read off that the fibration corresponds to turning on a dimension-two vector operator along the $x_3$ direction whose expectation value goes as $b^6/m$. 

Away from the UV limit, the field theory lives on the direct product of three-dimensional Minkowski spacetime and a circle. We can set $x_3\equiv \fft{m}{6b^4}\phi$, where $\phi$ has a period of $2\pi$. Then the circular direction has a physical radius of $\fft{m}{6b^4}$. Curiously enough, this radius can be made large enough to be consistent with observations by taking $b^4\ll m$. In fact, statistical tools to use the cosmic microwave background to search for ``circles in the sky'' have been developed in, for example, \cite{cornish}. If, on the other hand, the radius of the circular direction is small, then the field theory undergoes a flow ``across dimensions'' towards a three-dimensional theory.

The radius of the angular coordinate $\phi$ from the ten-dimensional viewpoint is given by $H^{-1/4} \fft{m}{6b^4}$, which decreases monotonically as one comes in from the asymptotic region. In order for the type IIB description to be reliable, this radius must be much larger than the string length scale. When this circle gets smaller than the string length scale, the more appropriate description is obtained by T-dualizing along the $x_3$ direction. Then one obtains a non-singular D2-brane solution given by \cite{lu1}
\bea\label{D2}
ds_{10}^2 &=& H^{-5/8} (-dt^2+dx_1^2+dx_2^2)+H^{3/8} (ds_6^2+dy^2),\\
F_{(4)} &=& dt\wedge dx_1\wedge dx_2\wedge dH^{-1}+\ast_6 L_{(2)},\qquad F_{(3)} = mL_{(2)}\wedge dy,\qquad \phi=-\fft14 \log H.\nn
\eea
Note that T-duality has untwisted the $x_3$, which now corresponds to the $y$ coordinate of the transverse space. 

A convenient method for determining the preserved supersymmetry of this solution is to now dimensionally oxidize it to eleven dimensions, which results in a non-singular M2-brane solution given by
\bea
ds_{11}^2 &=& H^{-2/3} (-dt^2+dx_1^2+dx_2^2)+H^{1/3} (ds_6^2+dy^2+dz^2),\nn\\
F_{(4)} &=& dt\wedge dx_1\wedge dx_2\wedge dH^{-1}+m L_{(4)},
\eea
where
\be
L_{(4)} = \ast_6 L_{(2)}+L_{(2)}\wedge dy\wedge dz,
\ee
is a self-dual harmonic 4-form on the 8-dimensional transverse space with the metric $ds_6^2+dy^2+dz^2$. The introduction of $L_{(4)}$ to the M2-brane solution preserves all of the initial supersymmetries, provided that \cite{becker}
\be
L_{abcd} \Gamma^{bcd}\epsilon =0,
\ee
where $\epsilon$ is a Killing spinor in the 8-dimensional transverse space. This implies that the $u_i$ must satisfy the linear relation:
\be
-u_0+u_1+u_2=0.
\ee
While the K\"ahler form does not satisfy this requirement, the second harmonic 2-form does and so the wrapped D3-brane solution is supersymmetric.

Alternatively, we can perform T-duality along the $\psi$ direction of the D3-brane solution, which untwists the $S^5/\Z_3$ to $\CP^2\times S^1$ \cite{untwisted}. Since $\CP^2$ does not admit a spin structure, its spectrum contains no fermions. While the resulting type IIA solution is not supersymmetric at the level of supergravity, the missing superpartners are provided by stringy winding modes. Performing T-duality along the $\psi$ direction and lifting to eleven dimensions yields the solution
\bea
ds_{11}^2 &=& \left( \fft{R}{3}\right)^{2/3} \left[ \fft{r^2}{R^2} \left( -dt^2+dx_1^2+dx_2^2+f dx_3^2\right)+\fft{R^2}{r^2} f^{-1} dr^2+R^2 ds_{\CP^2}^2+\fft{9}{R^2} (dy^2+dz^2)\right],\nn\\
F_\4 &=& \fft{2m}{3r} \left( \fft{1}{R^4}+\fft{1}{r^4}\right) dt\wedge dx_1\wedge dx_2\wedge dr
+ \fft{R^4}{3} \left( 4f-\fft{m^2}{r^{10}}\right) \epsilon_\4\nn\\
&+& \fft32 \left( \fft{mR^4}{18r^7} dx_3\wedge dr+dy\wedge dz\right)
\left( \sin^2\theta \sigma_1\wedge\sigma_2+\sin (2\theta) d\sigma_3\wedge d\theta\right),
\eea
where $\epsilon_\4$ is the volume-form of $\CP^2$. The geometry smoothly interpolates from AdS$_5\times \CP^2\times \R^2$ at large distance to Mink$_3\times \CP^2\times \R^4$ at short distance.

\subsection{On a resolved cone over $T^{1,1}/\Z_2$}

While there is an $S^1$-wrapped D3-brane solution on the resolved conifold whose geometry is asymptotically  AdS$_5\times T^{1,1}$, it has a naked singularity. On the other hand, there are $S^1$-wrapped D3-brane solutions on the resolved cone over $T^{1,1}/\Z_2$ which are asymptotically AdS$_5\times T^{1,1}/\Z_2$ and completely regular \cite{lu1}. The resolved cone over $T^{1,1}/\Z_2$ has the metric
\be\label{T11-cone}
ds_6^2=f^{-1} dr^2+\fft{r^2}{36} f (\sigma_3-\Sigma_3)^2+\fft{1}{12} R_1^2 (\sigma_1^2+\sigma_2^2)+\fft{1}{12} R_2^2 (\Sigma_1^2+\Sigma_2^2),
\ee
where $\sigma_i$ and $\Sigma_i$ are left-invariant 1-forms of $SU(2)\times SU(2)$, $R_i^2\equiv r^2+\ell_i^2$ and
\be
f=\fft{2r^4+3(\ell_1^2+\ell_2^2) r^2+6\ell_1^2 \ell_2^2}{R_1^2R_2^2}.
\ee
The radial coordinate $r\ge 0$. In order for the geometry described by the metric (\ref{T11-cone}) to be $\R^2\times S^2\times S^2$ at small distance for nonvanishing $\ell_1$ and $\ell_2$, the principal orbit must be $T^{1,1}/\Z_2$.

The metric (\ref{T11-cone}) was found in \cite{pando} with the radial coordinate $\rho$ and the parameters $a$ and $b$ given by
\be
2\rho^2=R_1^2,\qquad 12a^2=\ell_2^2-\ell_1^2,\qquad 16b^6=\ell_1^4 (3\ell_2^2-\ell_1^2),
\ee
where $a$ and $b$ correspond to a blown-up 2-cycle and a blown-up 4-cycle, respectively. This reduces to the metric of the resolved conifold \cite{Candelas} for vanishing $b$ and is included in the class of metrics obtained in \cite{berard,page} for vanishing $a$. 

A 2-form that preserves the isometry of (\ref{T11-cone}) has the form given by (\ref{L1}) expressed in the vielbein basis
\bea
e^0 &=& \fft{dr}{\sqrt{f}},\qquad e^1=\fft{R_1}{\sqrt{12}}\sigma_1,\qquad e^2=\fft{R_1}{\sqrt{12}}\sigma_2,\nn\\
e^3 &=& \fft{R_2}{\sqrt{12}}\Sigma_1,\qquad e^4=\fft{R_2}{\sqrt{12}}\Sigma_2,\qquad e^5=\fft{r}{6} \sqrt{f} (\sigma_3-\Sigma_3),
\eea
where the $u_i$ depend only on $r$. In this case, there are three harmonic 2-forms. Besides the K\"ahler form, there is one with
\be\label{T11-u1}
u_0 = \fft{(\ell_1^2-\ell_2^2)(r^4-\ell_1^2\ell_2^2)}{R_1^4 R_2^4},\qquad u_1=\fft{r^4+2\ell_1^2r^2+\ell_1^2 \ell_2^2}{R_1^4 R_2^2},\qquad u_2=\fft{r^4+2\ell_2^2r^2+\ell_1^2\ell_2^2}{R_1^2 R_2^4}.
\ee
This 2-form is square integrable at short distance. However, it is non-normalizable at large distance and carries non-trivial flux. The last harmonic 2-form is given by
\be\label{T11-u2}
u_0 = \fft{R_1^2+R_2^2}{R_1^4 R_2^4},\qquad u_1=\fft{1}{R_1^4 R_2^2},\qquad u_2=-\fft{1}{R_1^2 R_2^4}. 
\ee
This 2-form is square integrable at short distance and normalizable with vanishing flux. These two 2-forms can be used to have two of the D3-brane worldvolume directions fibered over the transverse space. This yields a regular solution with
\bea
H &=& 1+\fft{\ell_1^4 (\ell_1^2-\ell_2^2)^2 m_1^2+m_2^2}{4\ell_1^4 (\ell_1^2-3\ell_2^2)(\ell_1^2-\ell_2^2)R_1^2}+\fft{\ell_2^4 (\ell_1^2-\ell_2^2)^2 m_1^2+m_2^2}{4\ell_2^4 (\ell_2^2-3\ell_1^2)(\ell_2^2-\ell_1^2)R_2^2}\\
&+& \fft{\ell_1^4\ell_2^4 (\ell_1^2+\ell_2^2)^2 m_1^2+(\ell_1^2-\ell_2^2)^2 m_2^2}{2\sqrt{3} \ell_1^4\ell_2^4 (\ell_1^2-3\ell_2^2)^{3/2} (3\ell_1^2-\ell_2^2)^{3/2}} \log \left( 
\fft{4r^2+3(\ell_1^2+\ell_2^2)-\sqrt{3(\ell_1^2-3\ell_2^2)(3\ell_1^2-\ell_2^2)}}{4r^2+3(\ell_1^2+\ell_2^2)+\sqrt{3(\ell_1^2-3\ell_2^2)(3\ell_1^2-\ell_2^2)}}\nn
\right),
\eea
for $\ell_2\ne \ell_1$ and $\ell_2\ne \ell_1/\sqrt{3}$, where $m_1$ and $m_2$ are associated with the 2-forms with (\ref{T11-u1}) and (\ref{T11-u2}) and the wrapped worldvolume directions have periods $\pi m_1/3$ and $\pi m_2/(6\ell_1^2 \ell_2^2)$, respectively. For $\ell_2=\ell_1$ we have
\be
H=1+\fft{m_2^2}{8\ell_1^6 R_1^4}
+\fft{m_1^2}{2\sqrt{3} \ell_1^2} \arctan \left( \fft{\sqrt{3}\ell_1^2}{2r^2+3\ell_1^2}\right),
\ee
and for $\ell_2=\ell_1/\sqrt{3}$ we have
\be\label{D3-H1}
H=1+\fft{m_1^2\ell_1^6(27r^6+63\ell_1^2 r^4+45\ell_1^4 r^2+11\ell_1^6)+4m_2^2 (18r^4+36\ell_1^2r^2+19\ell_1^4)}{36\ell_1^6R_1^6 (3r^2+\ell_1^2)}.
\ee

One can also superimpose the 2-forms to construct a D3-brane solution with a single worldvolume direction fibered over the transverse space. The wrapped direction has a period of $\pi m_1/3$ and $\ell_1^2 \ell_2^2 m_1/m_2$ must be rational in order to have a smooth manifold. Note that $H$ may contain mixed terms in $m_1$ and $m_2$. For $\ell_2=\ell_1$, there are no mixed terms in $H$. On the other hand, for $\ell_2=\ell_1/\sqrt{3}$,
\bea\label{D3-H2}
H &=& 1+\fft{9m_1^2 (18r^4+36\ell_1^2 r^2+19\ell_1^4)+2\ell_1^6 m_2^2 (27r^6+63\ell_1^2 r^4+45\ell_1^4 r^2+11\ell_1^6)}{72\ell_1^6 R_1^6 (3r^2+\ell_1^2)}\nn\\
&+& \fft{m_1m_2 (9r^4+18\ell_1^2 r^2+11\ell_1^4)}{12\ell_1^2 R_1^6 (3r^2+\ell_1^2)}.
\eea
Note that (\ref{D3-H1}) does not arise in any limit of (\ref{D3-H2}), since the value of the integration constant that yields a regular solution itself depends on mixed terms in $m_1$ and $m_2$.

For vanishing $m_1$ and taking the integration constant in $H$ to be $0$ instead of $1$, $H$ has an asymptotic expansion which can be written as
\be
H\approx \fft{m_2^2 (\ell_1^2+\ell_2^2)}{64\ell_1^4 \ell_2^4 \rho^4} \left[ 1+\fft{6a^2}{\rho^2}+\cdots\right].
\ee
The geometry is AdS$_5\times T^{1,1}/\Z_2$ at large distance and a product space of three-dimensional Minkowski spacetime and a $U(1)$ bundle over $\R^2\times S^2\times S^2$ at short distance. The UV limit of the dual field theory is the $\Z_2$ orbifold of the conifold theory \cite{klebanov-witten} called the ${\mathbb F}_0$ theory \cite{morrison}. 

The blown-up 2-cycle, which is a global deformation in that it changes the position of the branes at infinity, corresponds to a dimension-two scalar operator that can be written in terms of the bifundamental fields as \cite{benvenuti}
\be
{\cal K}=A_{\alpha} \bar A^{\alpha}-B_{\dot\alpha}\bar B^{\dot\alpha}+C_{\alpha}\bar C^{\alpha}-D_{\dot\alpha} \bar D^{\dot\alpha}.
\ee
This operator lies within the $U(1)$ baryonic current multiplet, and its dimension is protected since the conserved current has no anomalous dimension. As one flows from the UV fixed point, this operator gets a vacuum expectation value that goes as $a^2$. The dimension-six non-mesonic scalar operator that corresponds to the blown-up 4-cycle gets a vacuum expectation value that goes as $b^6$. Also, the fibration corresponds to turning on a dimension-two vector operator along the $x_3$ direction with a vacuum expectation value that goes as $\fft{\ell_1^4\ell_2^4}{m_2(\ell_1^2+\ell_2^2)}$. 

We can perform T-duality along the $U(1)_R$ direction of $T^{1,1}/\Z_2$, which untwists $T^{1,1}/\Z_2$ to $S^2\times S^2\times S^1$ \cite{untwisted}. Although supersymmetry is broken at the level of supergravity, it is still preserved within the full string theory. Performing T-duality along the $\psi$ direction and lifting to eleven dimensions yields
\bea
ds_{11}^2 &=& h^{1/3} H^{-2/3} \Big[ -dt^2+dx_1^2+dx_2^2+H\Big( f^{-1} dr^2+h^{-1} \Big( \fft{fr^2}{36} dx_3^2+dy^2+dz^2\Big) \nn\\
&+& \fft{1}{12} R_1^2 (\sigma_1^2+\sigma_2^2)+\fft{1}{12} R_2^2 (\Sigma_1^2+\Sigma_2^2)\Big)\Big],\nn\\
F_\4 &=& dt\wedge dx_1\wedge dx_2\wedge \left( \fft{(R_1^2+R_2^2)r}{6R_1^4 R_2^4} dr-\fft{m}{12R_1^2 R_2^2} dH^{-1}\right)\nn\\
&+& \left[ \Sigma_1\wedge\Sigma_2-\sigma_1\wedge\sigma_2-d\left( \fft{m}{12hR_1^2 R_2^2}\right)\wedge dx_3\right]\wedge dy\wedge dz\nn\\
&+& \ast \Big[ dt\wedge dx_1\wedge dx_2\wedge dy\wedge dz\wedge \Big( \fft{Hfr^2}{36h} dx_3\wedge dH^{-1}-\fft{m(R_1^2+R_2^2)r}{72R_1^6 R_2^6 h} dx_3\wedge dr\nn\\
&+& \fft{1}{12R_1^2 R_2^2} (\sigma_1\wedge \sigma_2-\Sigma_1\wedge \Sigma_2)\Big)\Big],
\eea
where
\be
h=Hf\fft{r^2}{36}+\fft{m_2^2}{144R_1^4 R_2^4}>0.
\ee
Since $h>0$, this solution is free from curvature singularities. The geometry smoothly interpolates from AdS$_5\times S^2\times S^2\times \R^2$ at large distance to Mink$_3\times S^2\times S^2\times \R^4$ at short distance, where from now on we will use Mink$_d$ to refer to $d$-dimensional Minkowski spacetime. In order to avoid a conical singularity at $r=r_0$, the coordinate $x_3$ has a period of $\pi m_2/(6\ell_1^2\ell_2^2)$.

\subsection{On cones over $Y^{p,q}$}

We will consider cones over the $Y^{p,q}$ Sasaki-Einstein spaces  \cite{Gauntlett1,Gauntlett2} which have the metric
\be ds_6^2=K(r)^{-1}\,dr^2+\fft19 K(r)\,r^2\, \big(
d\psi+{\cal A}\big)^2
+r^2\,ds_4^2, \label{6cone} \ee
where
\be 
{\cal A}=-\cos\theta\,d\phi+y\,(d\beta+\cos\theta\,d\phi),\qquad K(r)=1-\fft{b^6}{r^6}, \label{K} \ee
and the radial coordinate $r\ge b$. The $Y^{p,q}$ metric has been expressed in canonical form with its base space given by the Einstein-K\"{a}hler metric
\be ds_4^2=\fft16 (1-y)(d\theta^2+\sin^2\theta\,d\phi^2)+
\fft{dy^2}{w(y)q(y)}+\fft{1}{36}w(y)q(y)\,(d\beta+\cos\theta\,d\phi)^2,
\label{ek4} \ee
with
\be w(y)=\fft{2(a-y^2)}{1-y},\qquad
q(y)=\fft{a-3y^2+2y^3}{a-y^2}. \ee
Note that $d{\cal A}=6J$, where $J$ is the K\"ahler form of the base space. The $S^2$ coordinates $\theta$ and $\phi$ have the ranges $0\le\theta < \pi$ and $0\le \phi < 2\pi$. 
The angular $y$ coordinate has the range $y_1\le y\le y_2$, where $y_i$ are the two smaller roots of
$q(y)$. $a$ and $y_i$ can be written in terms of the Chern numbers $p$ and $q$ as
\bea a &=& \fft12-\fft{p^2-3q^2}{4p^3}\,\sqrt{4p^2-3q^2},\nn\\
y_1 &=& \fft{1}{4p}\,\Big( 2p-3q-\sqrt{4p^2-3q^2}\Big),\nn\\
y_2 &=& \fft{1}{4p}\,\Big( 2p+3q-\sqrt{4p^2-3q^2}\Big).
\label{ay} \eea

The six-dimensional metric (\ref{6cone}) uses the general construction given in \cite{page} and has been considered in \cite{Pal,Sfetsos,benvenuti}\footnote{Cones over the larger family of $L^{pqr}$ Sasaki-Einstein spaces \cite{Cvetic1} have also been considered in \cite{Sfetsos}.}. While there is no curvature singularity for nonvanishing $b$, there are generally conifold fixed points of complex co-dimension two at the apex of the cone \cite{Sfetsos,benvenuti}. There are special cases for which these are orbifold fixed points and the string dynamics is well defined \cite{orbifold1,orbifold2}. Then the Sasaki-Einstein space is in the quasi-regular class and $\sqrt{4p^2-3q^2}$ is integer-valued. This implies that the volume of the $Y^{p,q}$ space is a rational fraction of the volume of the unit $S^5$, corresponding to rational central charges in the dual gauge theory
\cite{Gauntlett2}. The larger family of spaces with conifold rather than orbifold singularities corresponds to irrational central charges in the dual gauge theory. For the cases in which the singularities correspond to orbifold fixed points, we denote the cone as $C(Y^{p,q})/\Z_{N_1,N_2}$. This space becomes $S^2\times \C^2/\Z_{N_i}$ in the vicinity of the orbifold fixed-points at $r=r_0$, $y=y_i$. The simplest example of an orbifolded resolved cone is $C(Y^{1,1})/\Z_{4,2}$, for which $a=1$ and the $Y^{1,1}$ space is $S^5/\Z_2$. Examples for higher values of $p$ and $q$ include
\be C(Y^{7,3})/\Z_{70,56},\qquad C(Y^{7,5})/\Z_{126,84},\qquad
C(Y^{13,7})/\Z_{312,234},\qquad C(Y^{13,8})/\Z_{91,65}. \ee

One harmonic 2-form supported by these six-dimensional cones is given by
\be {\td L}_\2=\fft{1}{r^2\,(1-y)^2}\,(e^1\wedge e^2-e^3\wedge
e^4). \label{tdL2} \ee
This 2-form is square-integrable for $r\rightarrow b$ and the range in $y$ is such
that $L_\2$ never diverges. However, since $L_\2$ is not normalizable at large distance, the resulting ten-dimensional geometry will not be asymptotically AdS. Another harmonic 2-form is
\be L_\2=\fft{1}{r^6}\,(e^1\wedge e^2-2e^0\wedge e^5+e^3\wedge
e^4), \label{L2} \ee
which is square-integrable at short distance and normalizable at large distance. For the latter harmonic 2-form, we can easily find closed-form solutions for the function $H$ of the form $H=h(r)+g(y)$. However, in order for $\partial_y\,H$ to remain finite in the asymptotic region, we find that it must not depend on $y$. There is a solution given by
\be H=\fft{m^2}{4b^6\,r^4}. \label{D3H} \ee
The geometry interpolates from a product space of Mink$_3$ and $U(1)$ bundle over $C(Y^{p,q})$ at short distance to AdS$_5 \times Y^{p,q}$ for large distance. The dual field theory flows from an ${\cal N}=1$ superconformal quiver gauge theory \cite{Benvenuti} to an ${\cal N}=2$ three-dimensional theory. The fibration corresponds to a dimension-two vector operator along the $x_3$ direction, and the blown-up 4-cycle corresponds to a dimension-six non-mesonic scalar operator which gets a vacuum expectation value that goes as $b^6$.

Note that a completely regular cone over $Y^{2,1}$ has been constructed in \cite{Oota}, which involves a blown-up 2-cycle and 4-cycle. However, the harmonic 2-form supported by this cone, along with the function $H$, would necessarily have angular dependence.

\subsection{From lifting a five-dimensional magnetic AdS soliton}

Five-dimensional Einstein-Maxwell theory has a magnetically-charged AdS soliton solution given by
\bea\label{AdS-soliton}
ds_5^2 &=& g^2 r^2 \left[ -dt^2+dx_1^2+dx_2^2+f dx_3^2\right]+\fft{dr^2}{g^2r^2f},\nn\\
B_\1 &=& \fft{3gb^3}{r^2} dx_3,
\eea
where
\be
f=1-\fft{\mu}{r^4}-\fft{b^6}{r^6},
\ee
and $r\ge r_0$ where $r_0$ is the largest root of $f$. The circular $x_3$ direction smoothly caps off like a cigar geometry, provided that $x_3$ has a period of
\be
\Delta x_3=\frac{2\pi r_0^5}{g^2 (2\mu r_0^2+3b^6)}.
\ee
For vanishing $b$, this reduces to the AdS soliton \cite{Horowitz}. A multiple-charge generalization of this solution in five-dimensional ${\cal N}=2$ gauged $U(1)^3$ supergravity arises in a large mass limit of the global AdS solitons found in \cite{Ross}. However, we will focus on the equal-charge case of these solutions, for which the AdS soliton carries $U(1)_R$ graviphoton charge. Using the consistent Kaluza-Klein reduction ansatz presented in \cite{Buchel} then enables us to lift the solution (\ref{AdS-soliton}) to type IIB theory on a Sasaki-Einstein manifold $Y^{p,q}$. The resulting $S^1$-wrapped D3-brane solution can be expressed as
\bea
ds_{10}^2 &=& g^2 r^2 \left[ -dt^2+dx_1^2+dx_2^2+\td f (dx_3+A_\1)^2\right]+ \fft{1}{g^2 r^2} \left[ \fft{dr^2}{f}+\fft{r^2f}{9\td f} (d\psi+{\cal A})^2+r^2 ds_4^2\right],\nn\\
F_\5 &=& dt\wedge dx_1\wedge dx_2\wedge \left[ 4g^4 r^3 dx_3\wedge dr+\fft{2b^3}{g^2} J\right]+{\rm dual},
\eea
where 
\be
\td f=1-\fft{\mu}{r^4},\qquad A_\1=\fft{b^3}{3g^2r^4\td f} (d\psi+{\cal A}),
\ee
${\cal A}$ is given by (\ref{K}) and $J$ is the K\"ahler form of the base metric $ds_4^2$ given by (\ref{ek4}). For vanishing $\mu$, the transverse space reduces to the Ricci-flat cone given by (\ref{6cone}). For $b\neq 0$, $\td f>0$. This means that the period of $x_3$ is solely dictated by the connection on the fiber, and is therefore different from what is required in five dimensions. The period of $x_3$ is
\be
\Delta x_3=\fft{2\pi b^3}{3g^2 (r_0^2-\mu)}.
\ee

The asymptotic geometry is generally AdS$_5\times Y^{p,q}$. However, as discussed in the last subsection, using $Y^{p,q}$ spaces for this construction means that there are generally orbifold or conifold fixed points, with the exceptions of $S^5/\Z_3$ and $T^{1,1}/\Z_2$ for which the resulting ten-dimensional background is completely regular. 

One can also consider the AdS soliton in five-dimensional ${\cal N}=2$ gauged $U(1)^3$ supergravity with three independent charges \cite{Ross}. Using the consistent Kaluza-Klein reduction ansatz in \cite{Duff}, one can lift this solution on $S^5/\Z_3$ to obtain a more general D3-brane solution. This is something that cannot be done for $T^{1,1}/\Z_2$ \cite{Hoxha} or the $Y^{p,q}$ spaces in general \cite{Buchel}, since this would involve $SU(2)\times SU(2)$ vector multiplets, which it is inconsistent to retain in any truncation to the massless sector.

\subsection{On a six-dimensional Schwarzschild instanton}

A six-dimensional analog of the Schwarzschild instanton \cite{Hawking} has the metric
\be
ds_6^2=f^{-1} dr^2+f d\psi^2+\fft{r^2}{3} \left( d\Omega_2^2+d\td\Omega_2^2\right),
\ee
where
\be
f=1-\fft{r_0^3}{r^3},
\ee
the coordinate $\psi$ has a period of $4\pi r_0/3$ and $r\ge r_0$. Three harmonic 2-forms are given by
\be
L_\2^1=-\fft{3}{r^4} dr\wedge d\psi,\qquad L_\2^2=\Omega_\2,\qquad L_\2^3=\td\Omega_\2.
\ee
These can be used to construct a D3-brane wrapped on one, two or three directions which are fibered over the Schwarzschild instanton \cite{Nuts-bolts}. For the case of three wrapped directions, they must have the periods $4\pi m_1/(3r_0^2)$, $4\pi m_2$ and $4\pi m_3$, respectively, in order for the manifold to be simply connected. For the case in which the connections are used for a single fibered direction, the resulting manifold will be simply connected if $m_3=m_2=m_1/(3r_0^2)$. We can also have non-simply-connected smooth manifolds if the ratios $m_3/m_2$ and $3r_0^2 m_2/m_1$ are rational-valued. The corresponding function $H$ is given by
\be
H=1+\fft{m_1^2}{r_0^3r^3}+\fft{m_2^2+m_3^2}{2\sqrt{3}r_0^2} \left[ 2\arctan \left( \fft{2r+r_0}{\sqrt{3}r_0}\right) -\pi+\sqrt{3} \log \left( \fft{r^2+r_0 r+r_0^2}{r^2}\right)\right].
\ee

Since the circular $\psi$ direction has a radius that stabilizes, we can perform T-duality along this direction. Furthermore, if the fibration involves the $\psi$ direction, then the radius of $\psi$ is nowhere vanishing and the resulting T-dual solution can be completely regular. We will consider the case in which only $m_1$ is nonvanishing. Performing T-duality along the $\psi$ direction and lifting to eleven dimensions yields the solution
\bea
ds_{11}^2 &=& h^{\fft13} H^{-\fft23} \Big[ -dt^2+dx_1^2+dx_2^2 + H\Big( f^{-1} dr^2+\fft{r^2}{3} (d\Omega_2^2+d\td\Omega_2^2)+h^{-1} (f dx_3^2+ dy^2+dz^2)\Big)\Big],\nn\\
F_\4 &=& \fft{m^2 H^2}{27r_0^3 h} (f-fH^{-1}-1) \Omega_\2\wedge \td\Omega_\2
-\fft{m}{r^3} dt\wedge dx_1\wedge dx_2\wedge \left( dH^{-1}+\fft{3}{r} dr\right) \nn\\
&+& d\left( \fft{m}{r^3} h^{-1}\right)\wedge dx_3\wedge dy\wedge dz,
\eea
where
\be
h=f+\fft{m_1^2}{r_0^3r^3}>0.
\ee
Since $h>0$, this solution is free from curvature singularities.
In order to avoid a conical singularity at $r=r_0$, the coordinate $x_3$ has a period of $4\pi m_1/(3r_0^2)$.
Note that for $m_1=r_0^3$, $h=1$ and there is a Ricci-flat subspace that is the direct product of the six-dimensional Schwarzschild instanton and a 2-torus.

\subsection{On generalized Taub-NUT/BOLT instantons}

A six-dimensional generalization of the Taub-BOLT metric\footnote{Solutions based on more general Kerr-Taub-Bolt instantons have been constructed in \cite{Bena}.} is given by \cite{Awad}
\be\label{6D-bolt1}
ds_6^2=f^{-1} dr^2+4N^2 f d\td\psi^2+R^2 (d\Omega_2^2+d\td\Omega_2^2),
\ee
where 
\be
f=\fft{(r+N)(r-3N)}{3R^2},\qquad d\td\psi\equiv d\psi+\cos\theta d\phi+\cos\td\theta d\td\phi,\qquad R\equiv\sqrt{r^2-N^2},
\ee
$\psi$ has a period of $12\pi$ and $r\ge 3N$. The geometry goes from $\R^2\times S^2\times S^2$ at short distance to a $U(1)$ bundle over a cone over $S^2\times S^2$ at large distance. One could replace $S^2\times S^2$ by $\CP^2$, for which there is a regular Taub-NUT instanton as well, however we will focus on the $S^2\times S^2$ case. Note that this Taub-BOLT instanton does not admit a spin structure since each $S^2$ factor generates an element of $H_2$ of odd self-intersection \cite{Chamblin}, although it may still admit a $Spin^C$ structure.

Besides the Kahler form, this metric supports two harmonic 2-forms of the form (\ref{L1}) expressed in the vielbein basis
\bea
e^0 &=& \fft{dr}{\sqrt{f}},\qquad e^1=R d\theta,\qquad e^2=R \sin\theta d\phi,\nn\\
e^3 &=& R d\td\theta,\qquad e^4=R\sin\td\theta d\td\phi,\qquad e^5=\sqrt{f} d\td\psi,
\eea
with
\be
u_0=-\fft{4N}{(r\pm N)^3},\qquad u_1=u_2=\fft{r\pm 3N}{(r\pm N)^3},
\ee
which we will associate with $m_{\pm}$. Neither form is normalizable at large distance but they are both square integrable at short distance. Thus, we can use these two forms to construct a regular D3-brane wrapped on a 2-torus fibered over the Taub-BOLT instanton \cite{Nuts-bolts} with
\bea
H &=& 1-\fft{m_1^2 (11r^3+20N r^2-17N^2 r+14N^3)}{16N (r+N)^4}+\fft{4m_2^2}{N(r+N)}\nn\\
&+& \fft{m_2^2 (12r^2-3Nr+11N^2)}{6N(r-N)^3}+\fft{96m_2^2-11m_1^2}{32N^2} \log \left( \fft{r-N}{r+N}\right).
\eea
In order for the solution to be regular, the two 2-forms must be associated with different directions on the worldvolume the D3-brane. The wrapped directions have the periods $3\pi m_+/(2N)$ and $6\pi m_-/N$, where $2N$ must be integer-valued in order for the manifold to be regular.

An alternative six-dimensional generalization of the Taub-BOLT metric is given by\footnote{We could also consider the multiple nut parameter extension that found in \cite{Mann}.} \cite{page-pope3} 
\be\label{6D-bolt}
ds_6^2=f^{-1} dr^2+4N^2f d\td\psi^2+(r^2-N^2) d\Omega_2^2+r^2 d\td\Omega_2^2,
\ee
where
\be
f=\fft{(r+N)(r-2N)}{3r(r-N)},\qquad d\td\psi=d\psi+\cos\theta d\phi,
\ee
$\psi$ has a period of $4\pi$ and $r\ge 2N$. The geometry goes from $\R^2\times S^2\times S^2$ at short distance to a $U(1)$ bundle over a cone over $S^2\times S^2$ at large distance. Besides the Kahler form, this metric supports three harmonic 2-forms of the form (\ref{L1}) expressed in the vielbein basis
\bea
e^0 &=& \fft{dr}{\sqrt{f}},\qquad e^1=\sqrt{r^2-N^2} d\theta,\qquad e^2=\sqrt{r^2-N^2} \sin\theta d\phi,\nn\\
e^3 &=& r d\td\theta,\qquad e^4=r\sin\td\theta d\td\phi,\qquad e^5=2N\sqrt{f} d\td\psi. 
\eea
The three harmonic 2-forms have
\be\label{D3-form1}
u_0=\fft{3r^2-N^2}{r^2 (r^2-N^2)^2},\qquad u_1=\fft{2N}{r (r^2-N^2)^2},\qquad u_2=0,
\ee
\be\label{D3-form2}
u_0=\fft{r}{N(r^2-N^2)^2},\qquad u_1=\fft{3N^2-r^2}{2N^2(r^2-N^2)^2},\qquad u_2=0,
\ee
and
\be\label{D3-form3}
u_0=u_1=0,\qquad u_2=\fft{1}{N^2 r^2},
\ee
which we will associate with $m_1$ and $m_2$ and $m_3$, respectively. While the second and third forms have non-trivial flux and are not normalizable at large distance, all three of them are square integrable at short distance. Thus, we can use these three forms to construct a regular D3-brane wrapped on a 3-torus fibered over the Taub-BOLT instanton with
\bea
H &=& 1+\fft{1}{96N^5} \Bigg[ \fft{(4m_1^2+m_2^2)(3N^3+31N^2 r+29N r^2+9r^3)}{(r-N)(r+N)^3}+\fft{20m_1^2+41m_2^2+432m_3^2}{r+N}\nn\\
&-& \fft{144(m_1^2+m_3^2)}{r}+\fft{32(4m_1^2+m_2^2+18m_3^2)}{N} \log \left( \fft{r+N}{r}\right)\Bigg].
\eea
The wrapped directions have periods $4\pi m_1/(3N^2)$, $2\pi m_2/(3N^2)$ and $4\pi m_3/(N^2)$. Note that one can also use the superposition of 2-forms to construct a D3-brane solution with a single worldvolume direction fibered over the transverse space. For example, for the metric (\ref{6D-bolt}), this yields
\bea
H &=& 1-\fft{3(m_1^2+m_3^2)}{2N^5 r}+\fft{3(2m_1+m_2)^2}{32N^5 (r-N)}+\fft{(2m_1-m_2)^2 (r+3N)}{48N^4 (r+N)^3}\nn\\
&-& \fft{(10m_1-41m_2)(2m_1-m_2)+432m_3^2}{96N^5 (r+N)}+\fft{(2m_1-m_2)^2+18m_3^2}{3N^6} \log \left( \fft{r+N}{r}\right).
\eea
Regularity then requires that the ratios of $m_1$, $m_2$ and $m_3$ are all rational-valued.

Simply taking $r\rightarrow -r$ transforms the six-dimensional Taub-BOLT metric (\ref{6D-bolt}) into a Taub-NUT metric, where now $r\ge N$ \cite{page-pope3}. However, now there is only a single harmonic 2-form that is square integrable at short distance, providing a regular D3-brane wrapped on a circle that is fibered over the Taub-NUT instanton. 

Since the $\psi$ direction has a radius that stabilizes, we can T-dualize along this direction. As an example, consider the $T^2$-wrapped D3-brane on the Taub-BOLT instanton with the metric (\ref{6D-bolt}) and the 2-forms specified by (\ref{D3-form1}) and (\ref{D3-form2}). We will not consider the 2-form specified by (\ref{D3-form3}) since this one does not lie along the $\psi$ direction and can easily be added back in once we perform T-duality anyway. The 2-forms with (\ref{D3-form1}) and (\ref{D3-form2}) have the corresponding 1-form potentials
\bea
A^1_{(1)} &=& A_1 d\td\psi,\qquad A_1= -\frac{2Nm_1}{r(r^2-N^2)},\nn\\
A^2_{(1)} &=& A_2 d\td\psi,\qquad A_2=\frac{(r^2-3N^2)m_2}{2N^2(r^2-N^2)}.
\eea
We will associate $A^1_\1$ and $A^2_\1$ with the $x_1$ and $x_2$ directions, respectively.
After T-dualizing along the $\psi$ direction and lifting to eleven dimensions, we have
\bea\label{D3-11D}
ds_{11}^2 &=& h^{\frac{1}{3}} H^{-\frac{2}{3}}\Bigg[-dt^2+gh^{-1}\left(dx_1-g^{-1}A_1A_2 dx_2\right)^2+dx_3^2\nn\\
&+& H\left( f^{-1} dr^2+(r^2-N^2) d\Omega_2^2+r^2d\tilde{\Omega}_2^2+4N^2fg^{-1}dx_2^2
+h^{-1}(dy^2+dz^2)\right)\Bigg],\\
F_{(4)}&=& 
\ast \left( \td F_\5\wedge dy\wedge dz\right)
+ \left[ d(h^{-1} A_1)\wedge dx_1+d(h^{-1} A_2)\wedge dx_2-\Omega_\2\right]\wedge dy\wedge dz\nn\\
&+& dt\wedge dx_3\wedge \left( 2A_1A_2 \Omega_\2+dA_1\wedge dx_2+dA_2\wedge dx_1\right)
- 2NH^{\prime} f(r^2-N^2)r^2 \Omega_\2\wedge \td\Omega_\2,\nn
\eea
where
\bea
\td F_\5 &=& 2h^{-1} \Omega_\2\wedge \left( NH^{\prime} f(r^2-N^2)r^2\td\Omega_\2-A_1A_2 dt\wedge dx_3\right)\wedge (A_1 dx_1+A_2 dx_2)\nn\\
&+& dt\wedge dx_3\wedge \Omega_\2\wedge (A_2 dx_1+A_1 dx_2)+\fft{1}{2h} (dA_1^2-dA_2^2) \wedge d^4x.
\eea
and
\bea
g=4N^2Hf+A_2^2>0,\quad h=4N^2Hf+A_1^2+A_2^2>0.
\eea
For nonvanishing $m_1$ and $m_2$, a conical singularity at $r=2N$ is avoided provided that $x_2$ has a periodicity of $2\pi m_2/(3N^2)$ and the connection on the fiber implies that $x_1$ has a period of $4\pi m_1/(3N^2)$. If $m_2=0$, then the metric in (\ref{D3-11D}) reduces to
\bea\label{D3-11Dreduced}
ds_{11}^2 &=& h^{1/3} H^{-2/3} \Big[ -dt^2+dx_2^2+dx_3^2
+ H \Big( f^{-1} dr^2+(r^2-N^2) d\Omega_2^2+r^2 d\td\Omega_2^2\nn\\
&+& 4N^2 fh^{-1} dx_1^2+h^{-1} (dy^2+dz^2)\Big)\Big].
\eea
Now the $x_1$ direction must have a period of $4\pi m_1/(3N^2)$ in order to avoid a conical singularity at $r=2N$ and the period of $x_2$ is arbitrary. If instead $m_1=0$ then the metric reduces to (\ref{D3-11Dreduced}) with $m_1$ replaced by $m_2$, $x_1$ and $x_2$ interchanged and the periods adjusted accordingly.

\section{Toroidally-wrapped M2-branes}

The toroidally-wrapped M2-brane solution with maximum fibrations has the form \cite{chen}
\bea\label{wrappedm2}  
ds_{11}^2 &=& H^{-2/3} \eta_{\mu\nu} d\td x^{\mu} d\td x^{\nu}
+H^{1/3} ds_8^2,\nn\\
F_\4 &=& d^3\td x\wedge dH^{-1}
+\fft12 \epsilon_{\mu\nu\rho} d\td x^{\mu}\wedge d\td x^{\nu}\wedge dA_\1^{\rho}, 
\eea
where $d\td x^{\mu}\equiv dx^{\mu}+A_\1^{\mu}$, $dA_\1^{\mu}=m^{\mu}L_\2^{\mu}$ ($\mu$ not summed) are harmonic 2-forms in the transverse space of the metric $ds_8^2$, and $\mu,\nu=0,\dots 2$. The equations of motion
are satisfied if the function $H$ is given by
\be\label{H2} 
\square H=-\fft12 \sum_{\mu=0}^2 \left( m^{\mu} L_\2^{\mu}\right)^2,
\ee
where $\square$ is the Laplacian on $ds_8^2$.

Note that we can dimensionally reduce along the $x_2$ direction to get a fundamental string in type IIA theory given by
\bea\label{NS1}
ds_{10}^2 &=& H^{-3/4} \left( -(d\td x^0)^2+(d\td x^1)^2\right) +H^{1/4} ds_8^2,\nn\\
F_\4 &=& d\td x^0\wedge d\td x^1\wedge dA_\1^2,\nn\\
F_\3 &=& d\td x^0\wedge d\td x^1\wedge dH^{-1}+m d\td x^0\wedge L_\2^1-m d\td x^1\wedge L_\2^0,\nn\\
F_\2 &=& m L_\2^2,\\
e^{-2\phi} &=& H.\nn
\eea
$L_\2^0$ is associated with rotation, $L_\2^1$ is associated with a wrapped direction that is fibered over the transverse space, and $L_\2^2$ is associated with flux.

\subsection{On a regular cone over $S^7/\Z_4$}

A regular cone over $S^7/\Z_4$ is given by \cite{calabi}
\bea\label{cone2} 
ds_8^2 &=& f^{-1} dr^2+r^2 \Big( \fft{1}{9} f d\td\psi^2
+ d\sigma^2 +\fft{1}{4} \cos^2\sigma \sin^2\sigma \left( d\beta-\cos\theta_1 d\phi_1+\cos\theta_2 d\phi_2\right)^2\\
&+& \fft{1}{4} \cos^2\sigma (d\theta_1^2+\sin^2\theta_1 d\phi_1^2) + \fft{1}{4} \sin^2\sigma (d\theta_2^2+\sin^2\theta_2 d\phi_2^2)\Big),\nn 
\eea
where
\be\label{f-M2}
f=1-\fft{b^8}{r^8},\qquad d\td\psi\equiv d\psi+\fft34 \cos (2\sigma) d\beta-\fft32 \cos^2\sigma \cos\theta_1 d\phi_1 -\fft32 \sin^2\sigma \cos\theta_2 d\phi_2,
\ee
$\psi$ has a period of $3\pi/2$ and $r\ge b$. A 2-form that preserves the isometry of (\ref{cone2}) has the form
\bea\label{2form-M2}
L_{(2)}=u_0 e^0\wedge e^7+ u_1 e^1\wedge e^2+ u_2 e^3\wedge e^4
+ u_3 e^5\wedge e^6.
\eea
where we are using the vielbein
\bea 
e^0 &=& \fft{dr}{\sqrt{f}},\qquad e^1 = \fft{r}{2} \cos\sigma d\theta_1,\qquad e^2=\fft{r}{2} \cos\sigma \sin\theta_1 d\phi_1,\nn\\
e^3 &=& \fft{r}{2} \sin\sigma d\theta_2,\qquad e^4=\fft{r}{2} \sin\sigma \sin\theta_2 d\phi_2,\qquad 
e^5 = r d\sigma,\nn\\
e^6 &=& \fft{r}{2} \cos\sigma \sin\sigma \left( d\beta -\cos\theta_1 d\phi_1+\cos\theta_2 d\phi_2\right),\qquad
e^7 = \fft{r}{3} \sqrt{f} d\td\psi.\nn
\eea
Two harmonic 2-forms are specified by
\be
u_0=\fft{2}{r^6},\qquad u_1=u_2=u_3=\fft{1}{r^6},
\ee
and
\be
u_0=0,\qquad u_1=u_2=u_3=\fft{1}{r^2},
\ee
which we associate with $m_1$ and $m_2$ respectively. These 2-forms can be used to construct a regular M2-brane solution wrapped on a 2-torus which is fibered over the transverse space, for which
\be
H=1+\fft{7m_1^2}{8b^8 r^2}+\fft{3b^8 m_2^2-7m_1^2}{8b^{10}} \left[ \arctan \left( 1-\fft{\sqrt{2} r}{b}\right)+\arctan \left( 1+\fft{\sqrt{2} r}{b}\right)\right].
\ee
For $r\rightarrow \infty$,
\be
H\approx 1+\fft{3m_2^2}{8r^2}+\fft{7m_1^2-3m_2^2}{24 b^{10} r^6}+\cdots
\ee
In order for the solution to be regular, the two 2-forms must be associated with different directions on the worldvolume the M2-brane. The geometry is generally a direct product of Mink$_3$ and a cone over $S^7/\Z_4$ at large distance and Mink$_3\times \R^2\times \CP^3$ at short distance. However, for $m_2=0$ and setting the integration constant in $H$ to $0$ instead of $1$, the geometry is asymptotically AdS$_4\times S^7/\Z_4$. The UV limit of the dual theory has been conjectured to be a three-dimensional level $4$ $U(N)\times U(N)$ ${\cal N}=6$ superconformal Chern-Simons matter theory \cite{Aharony}. M2-branes (not wrapped) on the regular cone over $S^7/\Z_4$ have been considered in \cite{Singh,Krishnan}, where it was shown from the linearized form of the metric that the resolution of the cone corresponds to turning on a dimension-4 operator. For the above wrapped M2-brane solution, the fibration corresponds to turning on a dimension-zero vector operator whose expectation value goes as $b^{10}/m_1$.

\subsection{On a regular cone over $Q^{1,1,1}/\Z_2$}

The coset space $Q^{1,1,1}$ possesses $U(1)_R\times SU(2)^3$ symmetry \cite{DAuria}. The resolved cone over $Q^{1,1,1}/\Z_2$ has the metric \cite{cvetic1}
\be
ds_8^2=h^2 dr^2+\fft{r^2}{16h^2} \sigma^2+  \fft{1}{8} \sum_{i=1}^3 a_i^2 (d\Omega_2^i)^2,
\ee
where
\be
(d\Omega_2^i)^2=d\theta_i^2+\sin^2\theta_i d\phi_i^2,\qquad \sigma=d\psi+\sum_{i=1}^3 \cos\theta_i d\phi_i,
\ee
and
\be
h^2=\fft{3\prod_i a_i^2}{3r^6+4\sum_i \ell_i^2 r^4+6\sum_{i<j} \ell_i^2 \ell_j^2 r^2+12\prod_i \ell_i^2},\qquad a_i^2=r^2+\ell_i^2.
\ee
In order for the geometry to be $\R^2\times S^2\times S^2\times S^2$ at small distance as $r\rightarrow 0$ for nonvanishing $\ell_i$, $\psi$ must have a period of $2\pi$ and the principal orbit must be $Q^{1,1,1}/\Z_2$. 

While there are four harmonic 2-forms that live on this space, we will focus on the one that is square integrable at short distance and normalizable at large distance \cite{chen}, given by (\ref{2form-M2}) with
\be
e^0=h\ dr,\qquad e^7=c\sigma,\qquad e^{2j-1}=a_j d\theta_j,\qquad e^{2j}=a_j \sin\theta_j d\phi_j,
\ee
for $j=1,2,3$ and the functions
\be
u_0 = \fft{3r^4+2\sum_i \ell_i^2 r^2+\sum_{i\neq j} \ell_i^2 \ell_j^2}{\prod_i a_i^4},\quad 
u_1 = \fft{1}{a_1^4 a_2^2 a_3^2},\quad u_2 = \fft{1}{a_1^2 a_2^4 a_3^2},\quad u_3 = \fft{1}{a_1^2 a_2^2 a_3^4}.
\ee
For $\ell_i\equiv \ell$, a regular solution for $H$ can be written in closed-form as
\be\label{Q111-exactH}
H=\fft{m^2}{4\ell^8 (r^2+\ell^2)^3},
\ee
The geometry is AdS$_4\times Q^{1,1,1}/\Z_2$ at large distance and a product space of Mink$_2$ and a $U(1)$ bundle over $\R^2\times S^2\times S^2\times S^2$ at short distance. 

We can use perturbative methods to find regular solutions based on the exact solution given by (\ref{Q111-exactH}). In particular, for $\ell_1=\ell (1+\ep_1)$, $\ell_2=\ell (1+\ep_2)$ and $\ell_3=\ell$ with $\ep_1, \ep_2\ll 1$, we have
\bea
H &\approx& \fft{m^2}{4\ell^8 (r^2+\ell^2)^3}-\fft{(\ep_1+\ep_2) m^2 (4r^2+7\ell^2)}{6\ell^8 (r^2+\ell^2)^4}\\
&+& \fft{(\ep_1^2+\ep_2^2) m^2}{36\ell^{14}} \left[ \fft{\ell^2 (115\ell^8+95\ell^6 r^2-20\ell^4 r^4-48\ell^2 r^6-12r^8)}{(r^2+\ell^2)^5}+12\arctan \left( 1+\fft{2\ell^2}{r^2}\right)\right]\nn\\
&+& \fft{\ep_1 \ep_2 m^2 }{9\ell^{12}} \left[ \fft{50\ell^8+64\ell^6 r^2+32\ell^4 r^4+12\ell^2 r^6+3r^8}{(r^2+\ell^2)^5}-\fft{3}{\ell^2} \arctan \left( 1+\fft{2\ell^2}{r^2}\right)\right]+{\cal O}(\ep^3).\nn
\eea

We can also study the general properties of $H$ at short and large distance for arbitrary $\ell_i$ away from the above perturbative regime. At large distance,
\be
H\sim \fft{1}{r^6} \left( 1-\fft{\sum_i \ell_i^2}{r^2}+\cdots\right),
\ee
which is asymptotically AdS$_4\times Q^{1,1,1}/\Z_2$. At short distance,
\be
H\approx 1-\fft{m^2 \sum_{i\neq j\neq k} \ell_i^4 \ell_j^2 (\ell_j^2+2\ell_k^2)}{16 \prod_i \ell_i^8}\  r^2+\cdots,
\ee
from which we see that all of the $\ell_i$ must be nonvanishing in order for the solution to be regular. 

The UV limit of the dual field theory is a three-dimensional ${\cal N}=2$ superconformal field theory \cite{Fabbri}. More specifically, it has been conjectured that the dual field theory is a $U(N)^4$ Chern-Simons quiver gauge theory at level 2 \cite{Franco} (see also \cite{Amariti,Davey}).

\subsection{On a regular cone over the Stiefel manifold}

The Stiefel manifold $V_{5,2}$ is a homogeneous Sasaki-Einstein seven-manifold. The metric for a regular cone over $V_{5,2}$ was found in \cite{stenzel,pope1}. Since it has a non-collapsing 4-cycle, it supports a harmonic 4-form which is square integrable at short distance. However, it does not have a non-collapsing 2-cycle and therefore does not support a square-integrable harmonic 2-form. Here we consider a regular cone over $V_{5,2}/\Z_3$ using the general construction of \cite{page}, which does have a non-collapsing 2-cycle.  
Since $V_{5,2}$ is the coset manifold $SO(5)/SO(3)$, it will be useful to consider the left-invariant 1-forms $L_{AB}$ on the group manifold $SO(5)$, which  satisfy
\be
dL_{AB}=L_{AC}\wedge L_{CB},
\ee
for $A=1,\dots , 5$. We consider the $SO(3)$ subgroup by splitting the index $A=(1,2,i)$. The 1-forms in the coset $SO(5)/SO(3)$ are
\be
\sigma_i= L_{1i},\qquad \td\sigma_i= L_{2i},\qquad \nu = L_{12},
\ee
which obey
\bea
d\sigma_i &=& \nu\wedge \td\sigma_i+L_{ij}\wedge\sigma_j,\qquad d\td\sigma_i=-\nu\sigma\sigma_i+L_{ij}\wedge \td\sigma_j,\qquad d\nu=-\sigma_i\wedge\td\sigma_i,\nn\\
dL_{ij} &=& L_{ik}\wedge L_{kj}-\sigma_i\wedge\sigma_j-\td\sigma_i\wedge\td\sigma_j,
\eea
where the $L_{ij}$ are the left-invariant 1-forms for the $SO(3)$ subgroup.

A regular cone over $V_{5,2}$ is given by
\be\label{V52-cone}
ds_8^2=f^{-1} dr^2+\fft{9}{16} r^2 f \nu^2+\fft{3}{8} r^2 (\sigma_i^2+\td\sigma_i^2),
\ee
where
\be
f=1-\fft{b^8}{r^8}.
\ee
This supports a harmonic 2-form that is square integrable at short distance and normalizable at large distance, given by
\be
L_\2=\fft{6}{r^7} dr\wedge\nu+\fft{1}{r^6} \sigma_i\wedge\td\sigma_i.
\ee
A regular solution is given by
\be
H=\fft{m^2}{4b^8 r^6}.
\ee
An expression for the metric on $V_{5,2}$ in angular coordinates was found in \cite{bergman}. From this, it can be seen that regularity of the cone (\ref{V52-cone}) requires that the period of the coordinate associated with the $U(1)_R$ symmetry is reduced to one third of what it would otherwise be for the $V_{5,2}$ space. Thus, the asymptotic geometry is $AdS_4\times V_{5,2}/\Z_3$, whose dual field theory has been conjectured to be a three-dimensional ${\cal N}=2$ Chern-Simons-quiver theory with gauge group $U(N)_3\times U(N)_{-3}$ \cite{Martelli2}. Since the $SO(5)\times U(1)_R$ isometry of $V_{5,2}$ reduces to $SU(2)\times U(1)\times U(1)_R$ upon taking the quotient, this has supplied the first example of a {\it non-toric} AdS$_4/CFT_3$ duality.

\subsection{On a regular cone over $M^{3,2}$}

The coset space $M^{3,2}$ (also known as $M^{1,1,1}$) possesses $U(1)_R\times SU(3)\times SU(2)$ symmetry \cite{Witten}. A regular cone over $M^{3,2}$ is given by 
\be\label{metric-M32}
ds_8^2 = \fft{dr^2}{f}+\fft{a_1^2}{16} (\Sigma_1^2+\Sigma_2^2)+\fft{r^2 f}{16} \left( \Sigma_3+\ft32 \sin^2\mu\ \sigma_3\right)^2
+ \fft{3a_2^2}{32} \left[ 4d\mu^2+\sin^2\mu \left( \sigma_1^2+\sigma_2^2+\cos^2\mu \sigma_3^2\right)\right],
\ee
where
\be
f=\fft{r^6+\ft43 (\ell_1^2+2\ell_2^2)r^4+2\ell_2^2 (\ell_2^2+2\ell_1^2)r^2+4\ell_1^2\ell_2^4}{a_1^2 a_2^4},\qquad a_i^2=r^2+\ell_i^2.
\ee
and $\sigma_i$ and $\Sigma_i$ are left-invariant 1-forms on $SU(2)\times SU(2)$. A 2-form which preserves the isometry of (\ref{metric-M32}) has the form given by (\ref{2form-M2}), where we have chosen the vielbein basis as
\bea
e^0 &=& \fft{dr}{\sqrt{f}},\qquad e^1=\fft{a_1}{4}\Sigma_1,\qquad e^2=\fft{a_1}{4}\Sigma_2,\qquad e^3=\sqrt{\ft{3}{32}} a_2 \sin\mu \sigma_1,\qquad
e^4 = \sqrt{\ft{3}{32}} a_2 \sin\mu \sigma_2,\nn\\ 
e^5 &=& \sqrt{\ft38} a_2 d\mu,
\qquad e^6=\sqrt{\ft{3}{32}} a_2 \sin\mu \cos\mu \sigma_3,\qquad e^7=\fft{r\sqrt{f}}{4} (\Sigma_3+\ft32 \sin^2\mu \sigma_3).\nn
\eea

We will first consider the case of $\ell_i\equiv\ell$, for which there are solutions in closed form. It is simpler to transform the metric to the form
\be
ds_8^2= \fft{d\rho^2}{g}+\fft{\rho^2}{16} \left( \Sigma_1^2+\Sigma_2^2
+g\left( \Sigma_3+\ft32\sin^2\mu\ \sigma_3\right)^2
+ 6 d\mu^2+\ft32 \sin^2\mu ( \sigma_1^2+\sigma_2^2+\cos^2\mu\ \sigma_3^2)\right),
\ee
where
\be
g=1-\fft{b^8}{\rho^8},
\ee
and $\rho\ge b$. Transforming the vielbein appropriately, one harmonic 2-form is given by 
\be
u_0=0,\qquad u_1=\frac{2}{r^2},\qquad -u_2=u_3=\frac{1}{r^2}.
\ee
The corresponding regular solution for $H$ is
\bea
H=1+\fft{3m^2}{16b^2} \left[ \fft{\pi}{2}-\arctan \left( \fft{r^2}{b^2}\right)\right].
\eea
Another harmonic 2-form is
\be
u_0=\frac{1+\sqrt{13}}{2r^{4+2\sqrt{13}}},\,u_1=u_2=-u_3=\frac{1}{r^{4+2\sqrt{13}}},
\ee
and the corresponding $H$ is
\bea
H &=& c- \fft{(1+\sqrt{13})m^2}{64b^{6+4\sqrt{13}}} \left( 2\arctan \left[ 1-\fft{\sqrt{2}r}{b}\right] +2\arctan \left[ 1+\fft{\sqrt{2}r}{b}\right] +\log \left[ \fft{b^2-r^2}{b^2+r^2}\right]\right)\nn\\
&+& \fft{(1+\sqrt{13})m^2}{16(2\sqrt{13}-1)b^8 r^{4\sqrt{13}-2}}\ _2F_1 \left[ 1,\fft{1}{4}-\fft{\sqrt{13}}{2},\fft54-\fft{\sqrt{13}}{2},\fft{r^8}{b^8}\right],
\eea
where $_2F_1$ is the hypergeometric function and $c$ is an integration constant. It can be verified that $H$ is completely regular, going to a constant at small and large distance. For each case, the geometry is a direct product of Minowski$_3$ and a cone over $M^{3,2}$ at large distance and a product space of Minkowski$_2$ and a $U(1)$ bundle over $\R^2\times S^2\times \CP^2$ at short distance. A third harmonic 2-form has $u_i\sim r^{2\sqrt{13}-4}$ which does not correspond to any regular solutions.

Now we will consider the large and small-distance behavior of a regular solution for arbitrary $\ell_1$ and $\ell_2$. At large distance,
\be
u_0\approx \fft{1+\sqrt{13}}{2r^{4+2\sqrt{13}}}+\cdots,\qquad u_1\approx u_2\approx -u_3\approx \fft{1}{r^{4+2\sqrt{13}}}+\cdots,
\ee
and
\be
H\approx 1-\fft{0.03m^2}{r^{6+4\sqrt{13}}}+\cdots .
\ee
Thus, there are no values of the parameters $\ell_i$ for which the geometry is asymptotically AdS.
At short distance, $u_i$ goes to a finite constant provided that both $\ell_i$ vanish and
\be
H\sim 1-m^2 r^2+\cdots
\ee
where the precise coefficients depend on $\ell_i$ as well as the boundary conditions.

\subsection{On cones over Sasaki-Einstein spaces}

As we briefly outline here, one can consider eight-dimensional cones over the countably-infinite classes of Sasaki-Einstein spaces constructed in \cite{7D1,Cvetic1,7D2,7D3} which have a metric of the form
\be ds_8^2=K(r)^{-1}\,dr^2+K(r)\,r^2\, \big(d\psi+{\cal A}_\1\big)^2+r^2\,ds_6^2, \ee
where
\be K(r)=1-\fft{b^8}{r^8}, \ee
and the radial coordinate $r\ge b$. The Sasaki-Einstein spaces can be locally expressed in canonical form as a $U(1)$ bundle over an Einstein-K\"ahler base metric $ds_6^2$, where $d{\cal A}_\1$ is proportional to the associated K\"ahler form. While there is no curvature singularity for nonvanishing $b$, there are generally conifold or, in special cases, orbifold fixed points at the apex of the cone. 

Cones of this form support a harmonic 2-form $L_\2$ for which $L_\2^2\sim 1/r^{16}$, which implies that it is square-integrable at short distance and normalizable at large distance. There is a corresponding solution for the function $H\sim 1/r^{6}$, so that the geometry smoothly interpolates from a product space of Mink$_2$ and a $U(1)$ bundle over the cone at short distance to the direct product of AdS$_4$ and the Sasaki-Einstein space at large distance. In the UV limit, the dual field theories have been proposed to be three-dimensional supersymmetric Chern-Simons matter theories. Many such examples can be found, for instance, in \cite{Martelli, Hanany,Ueda}.

\subsection{On Schwarzschild instantons}

We will now consider wrapped M2-branes on Schwarzschild instantons. Note that a more general solution that has three types of M2-brane charges dissolved in the fluxes on the Schwarzschild instanton, as well as angular momentum, has been constructed in \cite{Bena}.

An eight-dimensional version of the Schwarzschild instanton \cite{Hawking} has the metric
\be
ds_8^2=f^{-1} dr^2+f d\psi^2+\fft{r^2}{5} \sum_{i=1}^3 (d\Omega_2^i)^2,
\ee
where
\be
f=1-\fft{r_0^5}{r^5},
\ee
the coordinate $\psi$ has a period of $4\pi r_0/5$ and $r\ge r_0$. There are four harmonic 2-forms given by
\be
L_\2^1=-\fft{5}{r^6} dr\wedge d\psi,\qquad
L_\2^2=\Omega_\2^1,\qquad L_\2^3=\Omega_\2^2,\qquad L_\2^4=\Omega_\2^3.
\ee
These can be used to construct an M2-brane wrapped on one or two directions which are fibered over the Schwarzschild instanton, for which
\bea
H &=& 1+\fft{m_1^2}{r_0^5 r^5}
+ \fft{m_2^2+m_3^2+m_4^2}{60r_0^2} \Bigg[ \sqrt{8(5+\sqrt{5})} \arctan \left( \fft{4r+(1+\sqrt{5})r_0}{r_0\sqrt{10-2\sqrt{5}}}\right)\nn\\
&-& \sqrt{8(5-\sqrt{5})} \arctan \left( \fft{4r+(1-\sqrt{5})r_0}{r_0\sqrt{10+2\sqrt{5}}}\right)
\nn\\
&+& \log \left( \fft{(r^4+r_0r^3+r_0^2r^2+r_0^3r+r_0^4)^4(r^2+\ft12 (1-\sqrt{5}) r_0r+r_0^2)^{1+\sqrt{5}}}{r^{20} (r^2+\ft12 (1+\sqrt{5}) r_0r+r_0^2)^{\sqrt{5}-1}}\right)\Bigg].
\eea

We can reduce along the $x_2$ direction to get a fundamental string in type IIA theory given by (\ref{NS1}). Since the $\psi$ direction has a radius that stabilizes, we can T-dualize along this direction. Furthermore, if the fundamental string solution has a fibration involving the $\psi$ direction, then the radius of $\psi$ is nowhere vanishing. This implies that the T-dual solution in type IIB theory is regular. As an example, consider an M2-brane wrapped on the $x_1$ direction using the harmonic 2-form $L_\2^1$, whose corresponding 1-form is
\be
A_\1=\fft{m_1}{r^5} d\psi.
\ee
Reducing to type IIA theory along $x_2$ and then T-dualizing along $\psi$ yields a type IIB pp-wave solution given by
\bea
ds_{10}^2 &=& h^{1/4} \left[ -H^{-1} dt^2+f^{-1} dr^2+\fft{r^2}{5} \sum_{i=1}^3 (d\Omega_2^i)^2 +h^{-1}\Big( f dx_1^2+H (dy+B_\1)^2\Big)\right],\nn\\
F_\3^{NS} &=& h^{-1} dt\wedge dx_1\wedge \left( H^{-1}f dH+\fft{5m_1^2}{r^{11}} dr\right),\nn\\ 
e^{-2\phi} &=& h,\nn
\eea
where
\bea
h &=& f+\fft{m_1^2}{r_0^5 r^5}>0,\nn\\
B_\1 &=& \left( \fft{m_1^3}{r_0^5 r^{10} H}+\fft{r_0^5}{m_1} \log H\right) dt.
\eea
A conical singularity at $r=r_0$ is avoided provided that $x_1$ has a periodicity of $4\pi m_1/(5r_0^4)$. For $m_1=r_0^5$, $h=1$ and the pp-wave propagates on a Schwarzschild instanton.

We can also consider a wrapped M2-brane on the direct product of two four-dimensional gravitational instantons. As an example, we consider the case involving two Schwarzschild instantons for which the metric of the transverse space is
\be
ds_8^2=f^{-1} dr^2+f d\psi^2+r^2 d\Omega_2^2+\td f^{-1} d\td r^2+\td f d\td\psi^2+\td r^2 d\td\Omega_2^2,
\ee
where
\be
f=1-\fft{r_0}{r},\qquad \td f=1-\fft{\td r_0}{\td r},
\ee
the coordinates $\psi$ and $\td\psi$ have periods $4\pi r_0$ and $4\pi \td r_0$, $r\ge r_0$ and $\td r\ge \td r_0$. We will consider fibrations involving the 1-forms
\be\label{1form-M2}
A_\1=\fft{m}{r} d\psi,\qquad \td A_\1=\fft{\td m}{\td r} d\td \psi,
\ee
which we will associate with the $x_1$ and $x_2$ directions, respectively. There is a regular solution with
\be\label{H-product}
H=1+\fft{m^2}{r_0 r}+\fft{\td m^2}{\td r_0 \td r}.
\ee
Note that $\psi$ and $\td\psi$ both have radii which stabilize and are nowhere vanishing. Thus, we can reduce along $\psi$ and T-dualize along $\td\psi$ to obtain the type IIB solution
\bea\label{T-M21}
ds_{10}^2 &=& (h\td h)^{1/4} \left[ -\fft{dt}{H}^2+\fft{dr^2}{f} +\fft{f}{h} dx_1^2+r^2 d\Omega_2^2+\fft{d\td r^2}{\td f}+ 
\fft{\td f}{ \td h} dx_2^2+\td r^2 d\td \Omega_2^2 + \fft{H}{h\td h} (dy+B_\1)^2\right],\nn\\
F_\5 &=& \fft{1}{h\td h}\ dt\wedge dx_1\wedge dx_2\wedge \left( \fft{m^2\td f}{r_0r^2} dr+\fft{\td m^2 f}{\td r_0\td r^2} d\td r\right)\wedge (dy+B_\1)+\mbox{dual},\nn\\
F_\3^{RR} &=& d\left( \fft{m}{rh}\right)\wedge dx_1\wedge (dy+B_\1)+\fft{\td m}{\td r h} dt\wedge dx_1\wedge \left[ Hf \left( dH^{-1}+\fft{d\td r}{\td r}\right)-\fft{m^2}{r^3}dr\right],\\
F_\3^{NS} &=& d\left( \fft{\td m}{\td r\td h}\right) \wedge dx_2\wedge (dy+B_\1)+\fft{m}{r\td h} dt\wedge dx_2\wedge \left[ H\td f \left( dH^{-1}+\fft{dr}{r}\right)-\fft{\td m^2}{\td r^3}d\td r\right],\nn\\
e^{2\phi} &=& h \td h^{-1},\nn
\eea
where
\be
h=Hf+\fft{m^2}{r^2}>0,\qquad \td h=H\td f+\fft{\td m^2}{\td r^2}>0,
\ee
and
\be
dB_\1=\fft{m\td m}{r\td r} \left( dH^{-1}+\fft{dr}{r}+\fft{d\td r}{\td r}\right)\wedge dt.
\ee
Conical singularities at $r=r_0$ and $\td r=\td r_0$ are avoided provided that $x_1$ and $x_2$ have periods $4\pi m$ and $4\pi \td m$, respectively.

Note that if we instead reduce along $\td\psi$ and T-dualize along $\psi$ then we obtain the S-dual of the solution (\ref{T-M21}). While one can also construct a regular solution with a single fibration from the 1-forms (\ref{1form-M2}), performing dimensional reduction and T-duality would then lead to a singular solution.

\section{Other brane solutions}

\subsection{Toroidally-wrapped D5-branes}

If the transverse space has a 2-cycle $L_\2$, then we can construct a regular toroidally-wrapped D5-brane, in which the circular directions are fibered over the transverse space \cite{overlapping}. The solution is given by
\bea\label{D5} 
ds_{10}^2 &=& H^{-1/4} \eta_{\mu\nu} d\td x^{\mu} d\td x^{\nu}+H^{3/4} ds_4^2,\nn\\
F_\3^{RR} &=& \ast_4 dH-\sum_{\mu=0}^5 dA_\1^{\mu}\wedge d\td x^{\mu},\nn\\ 
e^{-2\phi} &=& H,
\eea
where $d\td x^{\mu}=dx^{\mu}+A_\1^{\mu}$, $dA_\1^{\mu}=m^{\mu}L_\2^{\mu}$ ($\mu$ not summed) are harmonic 2-forms in the transverse space of the metric $ds_4^2$ and $\ast_4$ is the Hodge dual with respect to $ds_4^2$. The equations of motion are satisfied, provided that
\be
\square H = -\fft12 \sum_{\mu=0}^5 \left( m^{\mu} L_\2^{\mu}\right)^2,
\ee
where $\square$ is the Laplacian on $ds_4^2$.

\subsubsection{On an Eguchi-Hanson instanton}

The metric for the Eguchi-Hanson is \cite{Eguchi}
\be
ds_4^2=f^{-1} dr^2+\fft14 r^2 f \sigma_3^2+\fft14 r^2 (\sigma_1^2+\sigma_2^2),
\ee
where
\be
f=1-\fft{a^4}{r^4},
\ee
and $\sigma_i$ are left-invariant 1-forms on $SU(2)$. The radial coordinate $r\ge a$ and the period of $\psi$ is $2\pi$ so that the principal orbit is $S^3/\Z_2$ in order to ensure regularity at $r=a$. The metric admits a harmonic 2-form that is square integrable at short distance, given by
\be
L_\2=\fft{1}{r^3} dr\wedge \sigma_3+\fft{1}{2r^2} \sigma_1\wedge\sigma_2.
\ee
We can use this to construct a regular D5-brane on a circle which is fibered over the Eguchi-Hanson instanton, for which
\be
H=1+\fft{m^2}{a^4 r^2}.
\ee
The connection on the fiber implies that the wrapped direction has a period of $\pi m/a^2$. The geometry is asymptotically Mink$_6\times \R^4/\Z_2$ and a product space of Mink$_5$ and a $U(1)$ bundle over $\R^2\times S^2$ at short distance

\subsubsection{On a Schwarzschild instanton}

The metric for the Schwarzschild instanton is \cite{Hawking}
\be
ds_4^2 = f^{-1} dr^2+fd\psi^2+r^2 d\Omega_2^2,
\ee
where
\be
f=1-\fft{r_0}{r},
\ee
and $r\ge r_0$. In order to avoid a conical singularity at $r=r_0$, the coordinate $\psi$ has a periodicity of $4\pi r_0$. This supports two harmonic 2-forms given by
\be
L_\2^1=-\fft{1}{r^2} dr\wedge d\psi,\qquad L_\2^2=\Omega_\2.
\ee
These can be used to construct a D5-brane wrapped on one or two fibered directions with
\be
H=1+\fft{m_1^2+m_2^2}{r_0r}.
\ee
For the case of two fibered directions, the connections on the fibers associated with $L_\2^1$ and $L_\2^2$ imply that the directions have periods $4\pi m_1$ and $4\pi m_2$, respectively. For the case in which both connections are used for a single fibered direction, the resulting manifold will be simply connected if $m_1=m_2$. We can also have non-simply-connected smooth manifolds if the ratio $m_1/m_2$ is rational-valued.

Since the circular $\psi$ direction has a radius that stabilizes, we can perform T-duality along this direction. Furthermore, if the fibration involves the $\psi$ direction, then the radius of $\psi$ is nowhere vanishing and the resulting type IIA solution can be completely regular. Consider an $S^1$-wrapped D5-brane for which only $L_\2^1$ is turned on, which corresponds to the 1-form
\be\label{1-form-D5}
A_\1=\fft{m}{r} d\psi.
\ee
Upon T-dualizing along the $\psi$ direction and performing dimensional oxidation, we obtain a regular solution in eleven dimensions given by
\bea
ds_{11}^2 &=& h^{1/3} H^{-1/3} \Big[ -dt^2+dx_1^2+\cdots dx_4^2+h^{-1} \left(dz+B_\1\right)^2 \nn\\
&+& H (f^{-1} dr^2+r^2 d\Omega_2^2+fh^{-1} dx_5^2+h^{-1} dy^2)\Big],\\
F_\4 &=& \fft{m^3 f}{hr_0r} \Omega_\2\wedge dx_5\wedge dy+d\left( \fft{m}{hr}\right) \wedge dx_5\wedge dy\wedge \left( dz+B_\1\right) ,\nn
\eea
where
\be
h=f+\fft{m^2}{r_0 r}>0,\qquad B_\1=\fft{m^2f}{r_0}\cos\theta d\phi+\fft{m}{r} dx_5.
\ee
This is an $S^1$-wrapped M5-brane solution, for which $x_5$ is a coordinate on the transverse space while $z$ is a worldvolume coordinate fibered over the transverse space. Since $h>0$, this solution is free from curvature singularities. A conical singularity at $r=r_0$ is avoided provided that $x_5$ has a periodicity of $4\pi m$. Then the connection on the fiber implies that $z$ has a period of $4\pi m^2/r_0$. The transverse space is not generally Ricci-flat due to the presence of the $h$ function. However, for $m=r_0$, $h=1$ and the transverse space reduces to the direct product of the Schwarzschild instanton and a circle.

\subsubsection{On Taub-NUT and Taub-BOLT instantons}

The Taub-NUT/BOLT instanton \cite{Taub} has the metric
\be
ds_4^2=f^{-1} dr^2+4N^2f (d\psi- \cos\theta d\phi)^2+(r^2-N^2) d\Omega_2^2,
\ee
where
\be
f=\fft{r^2-2Mr+N^2}{r^2-N^2},\qquad d\Omega_2^2=d\theta^2+\sin^2\theta d\phi^2.
\ee
For the Taub-NUT instanton $M=N$ and $r\ge N$, while for the Taub-BOLT instanton $M=\ft54 N$ and $r\ge 2N$. In both cases, in order to avoid a conical singularity and for the principal orbits to be regular, $\psi$ must have a period of $4\pi$.

Harmonic 2-forms supported by this metric are given by
\be\label{Lpm}
L_\2^{\pm}=\fft{1}{(r\pm N)^2} \left[ 2N dr\wedge (d\psi- \cos\theta d\phi) \pm (r^2-N^2) \Omega_\2\right] ,
\ee
where $\Omega_\2$ is the volume-form corresponding to $d\Omega_2^2$. The 1-form potentials are
\be
A_\1^{\pm} = \pm A_{\pm} (d\psi-\cos\theta d\phi),\qquad A_{\pm}=m_{\pm} \left( \fft{r\mp N}{r\pm N}\right).
\ee
For the Taub-NUT instanton, only $L_\2^+$ yields a regular solution with a single fibration, for which
\be
H=1+\fft{m_+^2}{4N(r+N)}.
\ee
For the Taub-BOLT instanton, both $L_\2^+$ and $L_\2^-$ are square integrable for $r\rightarrow 2N$. Therefore, 
these 2-forms can be used to construct a regular D5-brane solution wrapped on a 2-torus which is fibered over the transverse space, for which $H$ is given by
\be\label{D5-H}
H=1+\fft{4m_+^2}{9N(r+N)}-\fft{4m_-^2}{N(r-N)}.
\ee
The connections $A_\1^+$ and $A_\1^-$ on the fibers imply that the wrapped directions have periods $4\pi m_+/3$ and $12\pi m_-$, respectively. Alternatively, we can superimpose the 2-forms to construct a solution with a single worldvolume direction fibered over the transverse space, for which $H$ is again given by (\ref{D5-H}). Then the resulting manifold is simply connected if $m_+=9m_-$ and non-simply connected though still smooth if the ratio $m_+/m_-$ is rational-valued.

Since the circular $\psi$ direction has a radius that stabilizes, we can perform T-duality along this direction. We will consider the case in which the D5-brane is wrapped on a 2-torus which is fibered over the Taub-BOLT space, for which $A_\1^+$ and $A_\1^-$ are associated with the $x_4$ and $x_5$ directions, respectively. Upon T-dualizing along the $\psi$ direction and lifting to eleven dimensions, we get the solution
\bea\label{D5-11D}
ds_{11}^2 &=& h^{1/3} H^{-1/3} \Big[ -dt^2+dx_1^2+dx_2^2+dx_3^2+gh^{-1} \left( dx_4+m_+ m_- g^{-1} dx_5\right)^2+h^{-1} (dz+B_\1)^2\nn\\
&+& H\left( f^{-1} dr^2+(r^2-N^2) d\Omega_2^2+4N^2 fg^{-1} dx_5^2+h^{-1} dy^2\right) \Big],\nn\\
F_\4 &=& 
\left[ 1+h^{-1} \left( 2N(r^2-N^2)H^{\prime}f-A_+^2-A_-^2\right)\right] (A_- dx_5-A_+ dx_4)\wedge \Omega_\2\wedge dy\nn\\
&+& \left[ \Omega_\2+ d\left( h^{-1}A_+\right)\wedge dx_4-d\left( h^{-1}A_-\right)\wedge dx_5 \right] \wedge dy\wedge (dz+B_\1)\nn\\
&+& \fft{2Nm_+ m_-}{h (r^2-N^2)} dr\wedge dx_4\wedge dx_5\wedge dy,
\nn
\eea
where
\be
g = 4N^2 Hf+A_-^2>0,\qquad h = 4N^2 Hf+A_+^2+A_-^2>0,
\ee
and
\be
dB_\1= \left[ 2N(r^2-N^2) fH^{\prime}-A_+^2-A_-^2\right]  \Omega_\2+2N dr\wedge \left( \fft{m_+}{(r+N)^2} dx_4+\fft{m_-}{(r-N)^2} dx_5\right).
\ee
This solution describes two $T^2$-wrapped overlapping M5-branes. Since $g>0$ and $h>0$, this solution is free from curvature singularities. 

For nonvanishing $m_\pm$, a conical singularity at $r=2N$ is avoided provided that $x_5$ has a periodicity of $12\pi m_-$. Furthermore, the connections on the fibers imply that $x_4$ has a period of $4\pi m_+/3$ and that the period of $z$ is given by
\be
\Delta z= \fft{4\pi (m_+^2+81m_-^2)}{9n_1}=\fft{4\pi m_+^2}{3n_2}=\fft{24\pi m_-^2}{n_3},
\ee
where $n_1$, $n_2$ and $n_3$ are integers. An example of a non-simply connected manifold is provided by $n_1=4$, $n_2=3$ and $n_3=2$.

If $m_-=0$, then the metric in (\ref{D5-11D}) reduces to
\bea\label{D5-11Dreduced}
ds_{11}^2 &=& h^{1/3} H^{-1/3} \Big[ -dt^2+dx_1^2+dx_2^2+dx_3^2+dx_5^2+h^{-1} (dz+B_\1)^2\nn\\
&+& H \left( f^{-1} dr^2+(r^2-N^2) d\Omega_2^2+4N^2fh^{-1} dx_4^2+h^{-1} dy^2\right].
\eea
Now the $x_4$ direction must have a period of $4\pi m_+/3$ in order to avoid a conical singularity at $r=2N$ and the period of $x_5$ is arbitrary. The connection on the fiber implies that $z$ can have a period of $4\pi m_+^2/9$. On the other hand, if $m_+=0$ then the metric reduces to (\ref{D5-11Dreduced}) with $m_-$ replaced by $m_+$, $x_4$ and $x_5$ interchanged and the periods adjusted accordingly.

\subsection{More branes on Schwarzschild instantons}

We consider the $n$-dimensional version of the Schwarzschild instanton \cite{Hawking} with the metric
\be\label{n-metric}
ds_n^2=f^{-1} dr^2+fd\psi^2+r^2 d\Omega_{n-2}^2,
\ee
where
\be
f=1-\fft{r_0^{n-3}}{r^{n-3}},
\ee
the coordinate $\psi$ has a period of $4\pi r_0/(n-3)$ and $r\ge r_0$. This supports the harmonic 2-form
\be\label{n-form}
L_\2=-\fft{n-3}{r^{n-2}} dr\wedge d\psi,\qquad A_\1=\fft{m}{r^{n-3}} d\psi,
\ee
This yields a regular solution to
\be\label{H-eqn}
\square H=-\fft12 m^2 L_\2^2,
\ee
given by
\be\label{n-H}
H=1+\fft{m^2}{(r_0 r)^{n-3}}.
\ee

\subsubsection{$S^1$-wrapped M5-brane}

An $S^1$-wrapped M5-brane on a five-dimensional Schwarzschild instanton is given by
\bea
ds_{11}^2 &=& H^{-1/3} \left( -dt^2+dx_1^2+\cdots +dx_4^2+(dx_5+A_\1)^2\right) +H^{2/3} ds_5^2,\nn\\
F_\4 &=& \ast_5 dH-m\left( \ast_5 L_\2\right) \wedge (dx_5+A_\1),
\eea
where the Schwarzschild instanton metric is given by (\ref{n-metric}) with $n=5$, $H$ satisfies (\ref{H-eqn}), and $dA_\1=mL_2$. A regular solution has the function $H$ given by (\ref{n-H}) with $n=5$. The wrapped direction $x_5$ has the period $2\pi m/r_0$.

Performing dimensional reduction along the $x_5$ direction yields a D4-brane in type IIA theory whose resolution is associated with flux \cite{transgression}
\bea
ds_{10}^2 &=& H^{-3/8} (-dt^2+dx_1^2+\cdots + dx_4^2)+H^{5/8} ds_5^2,\nn\\
F_\4 &=& \ast_5 dH,\qquad F_\3 = m\ast_5 L_\2,\qquad F_\2 = mL_\2,\\
e^{-4\phi} &=& H.\nn
\eea
Alternatively, if $x_4$ is the fibered direction then reducing along $x_5$ yields an $S^1$-wrapped D4-brane. T-dualizing along $\psi$ yields a type IIB solution given by
\bea
ds_{10}^2 &=& h^{1/4} \left[ H^{-1/2} (-dt^2+dx_1^2+dx_2^2+dx_3^2)+H^{1/2} (f^{-1} dr^2+r^2 d\Omega_3^2+fh^{-1} dx_4^2+h^{-1} dy^2)\right],\nn\\
F_\5 &=& 2mfh^{-1} \Omega_\3 \wedge dx_4\wedge dy+\fft{r_0^2}{m} dt\wedge dx_1\wedge dx_2\wedge dx_3\wedge dH^{-1},\nn\\
F_\3^{RR} &=& \fft{2m^2}{r_0^2} \Omega_\3,\qquad F_\3^{NS} = d\left( \fft{m}{r^2h}\right)\wedge dx_4\wedge dy,\\
e^{-2\phi} &=& h.\nn
\eea
where
\be
h=f+\fft{m^2}{r_0^2 r^2}>0.
\ee
This describes a D3-brane wrapped on a circle. For $m=r_0^2$, $h=1$ and the transverse space reduces to a direct product of a five-dimensional Schwarzschild instanton and a circle. In order to avoid a conical singularity at $r=r_0$, $x_4$ has the period $2\pi m/r_0$.

\subsubsection{$S^1$-wrapped D2-brane}

An $S^1$-wrapped D2-brane on a seven-dimensional Schwarzschild instanton in type IIA theory is given by
\bea
ds_{10}^2 &=& H^{-5/8} \left( -dt^2+dx_1^2+(dx_2+A_\1)^2\right) +H^{3/8} ds_7^2,\nn\\
F_\4 &=& dt\wedge dx_1\wedge (dx_2+A_\1)\wedge dH^{-1}+m\ dt\wedge dx_1\wedge L_\2,\nn\\
e^{4\phi} &=& H,
\eea
where the Schwarzschild instanton metric is given by (\ref{n-metric}) with $n=7$, $H$ satisfies (\ref{H-eqn}), and $dA_\1=mL_2$. A regular solution has the function $H$ given by (\ref{n-H}) with $n=7$. The wrapped direction $x_2$ has the period $\pi m/r_0^3$.

T-dualizing along $\psi$ yields the type IIB solution
\bea
ds_{10}^2 &=& h^{1/4} \left[ H^{-3/4} (-dt^2+dx_1^2)+H^{1/4} (f^{-1} dr^2+r^2 d\Omega_5^2+fh^{-1} dx_2^2+h^{-1} dy^2)\right],\nn\\
F_\5 &=& \fft{4m^2}{r_0^4 r^4} (r_0^4 H-r^4 f)\left( \fft{1}{r^4 Hh} dt\wedge dx_1\wedge dx_2\wedge dy\wedge dr+\Omega_\5\right),\nn\\
F_\3^{RR} &=& \fft{r_0^4}{m} (1-H-H^2) dt\wedge dx_1\wedge dH^{-1},\qquad
F_\3^{NS} = d\left( \fft{m}{r^4h}\right)\wedge dx_2\wedge dy,\\
e^{2\phi} &=& Hh^{-1},\nn
\eea
where
\be
h=f+\fft{m^2}{r_0^4 r^4}>0.
\ee
This describes a D1-brane wrapped on a circle. For $m=r_0^4$, $h=1$ and the transverse space reduces to the direct product of a seven-dimensional Schwarzschild instanton and a circle. In order to avoid a conical singularity at $r=r_0$, $x_2$ has the period $\pi m/r_0^3$.

\subsubsection{$S^1$-wrapped D1-brane}

An $S^1$-wrapped D1-brane on an eight-dimensional Schwarzschild instanton in type IIB theory is given by
\bea
ds_{10}^2 &=& H^{-3/4} \left( -dt^2+(dx+A_\1)^2\right) +H^{1/4} ds_8^2,\nn\\
F_\3^{RR} &=& dt\wedge (dx_1+A_\1)\wedge dH^{-1}+m\ dt\wedge L_\2,\nn\\
e^{2\phi} &=& H,
\eea
where the Schwarzschild instanton metric is given by (\ref{n-metric}) with $n=8$, $H$ satisfies (\ref{H-eqn}), and $dA_\1=mL_2$. A regular solution has the function $H$ given by (\ref{n-H}) with $n=8$. The wrapped direction $x$ has the period $4\pi m/(5r_0^4)$.

T-dualizing along $\psi$ and lifting to eleven dimensions yields the pp-wave solution
\bea
ds_{11}^2 &=& h^{1/3} \Big[ -H^{-1} dt^2+f^{-1} dr^2+r^2 d\Omega_6^2+h^{-1} \Big( f dx^2+dy^2+H (dz-B_\1)^2\Big) \Big],\nn\\
F_\4 &=& h^{-1} (fH^{-1}+f-1) dt\wedge dx\wedge dy\wedge dH+d\left( \fft{m}{r^5h}\right)\wedge dx\wedge dy\wedge dz,
\eea
where
\be
h=f+\fft{m^2}{r_0^5 r^5}>0,\qquad dB_\1=\fft{m}{r^5} dt\wedge dH^{-1}+\fft{r_0^5}{m} dt\wedge dH.
\ee
In order to avoid a conical singularity at $r=r_0$, $x$ has the period $4\pi m/(5r_0^4)$. For $m=r_0^5$, $h=1$ and the pp-wave propagates on a Schwarzschild instanton.

One can also consider a wrapped D1-brane on the direct product of two four-dimensional Schwarzschild instantons. However, in this case, the T-dual solution is not regular.

\subsubsection{Rotating D0-brane}

A rotating D0-brane on a nine-dimensional Schwarzschild instanton in type IIA theory is given by
\bea
ds_{10}^2 &=& -H^{-7/8} (dt+A_\1)^2+H^{1/8} ds_9^2,\nn\\
F_\2 &=& (dt+A_\1)\wedge dH^{-1}+m L_\2,\nn\\
e^{\fft43 \phi} &=& H,
\eea
where the Schwarzschild instanton metric is given by (\ref{n-metric}) with $n=9$, $H$ satisfies (\ref{H-eqn}), and $dA_\1=mL_2$. A regular solution has the function $H$ given by (\ref{n-H}) with $n=9$. Although the $\psi$ direction has a radius that stabilizes, it also vanishes for a certain value of $r$. If one T-dualizes over $\psi$, then this would lead to a singular solution.

\section*{Acknowledgments}

We are grateful to Philip Argyres, Mohammad Edalati, Gary Horowitz, Hong L\"u and Leopoldo Pando Zayas for useful discussions. This research was supported in part by the National Science Foundation under Grant Nos. NSF PHY-0969482 and NSF PHY11-25915, and a PSC-CUNY Award jointly funded by The Professional Staff Congress and The City University of New York.


\begin{thebibliography}{99}

\bibitem{agmoo} 
  O.~Aharony, S.~S.~Gubser, J.~M.~Maldacena, H.~Ooguri and Y.~Oz,
  ``Large N field theories, string theory and gravity,''
  Phys.\ Rept.\  {\bf 323}, 183 (2000)
  [hep-th/9905111].

\bibitem{KS} 
  I.~R.~Klebanov and M.~J.~Strassler,
  ``Supergravity and a confining gauge theory: Duality cascades and $\chi$SB resolution of naked singularities,''
  JHEP {\bf 0008}, 052 (2000)
  [hep-th/0007191].

\bibitem{transgression}
  M.~Cveti\v{c}, H.~L\"u, C.~N.~Pope,
  ``Brane resolution through transgression,''
  Nucl.\ Phys.\  {\bf B600}, 103-132 (2001).
  [hep-th/0011023].
    
\bibitem{pope1}
  M.~Cveti\v{c}, G.~W.~Gibbons, H.~L\"u, C.~N.~Pope,
  ``Ricci flat metrics, harmonic forms and brane resolutions,''
  Commun.\ Math.\ Phys.\  {\bf 232}, 457-500 (2003).
  [hep-th/0012011].

\bibitem{cvetic1}
  M.~Cveti\v{c}, G.~W.~Gibbons, H.~L\"u, C.~N.~Pope,
  ``Supersymmetric nonsingular fractional D2-branes and NS-NS 2-branes,''
  Nucl.\ Phys.\  {\bf B606}, 18-44 (2001).
  [hep-th/0101096].

\bibitem{overlapping} 
  H.~L\"u and J.~F.~V\'azquez-Poritz,
  ``Resolution of overlapping branes,''
  Phys.\ Lett.\ B {\bf 534}, 155 (2002)
  [hep-th/0202075].

\bibitem{lu1}
  H.~L\"u, J.~F.~V\'azquez-Poritz,
  ``$S^1$-wrapped D3-branes on conifolds,''
  Nucl.\ Phys.\  {\bf B633}, 114-126 (2002).
  [hep-th/0202175].

\bibitem{chen}
  C.~-M.~Chen, J.~F.~V\'azquez-Poritz,
  ``Resolving the M2-brane,''
  Class.\ Quant.\ Grav.\  {\bf 22}, 4231-4246 (2005).
  [hep-th/0403109].
 
\bibitem{Nuts-bolts} 
  J.~F.~V\'azquez-Poritz,
  ``The Nuts and bolts of brane resolution,''
  hep-th/0408144.
            
\bibitem{kachru}
  S.~Kachru and E.~Silverstein,
  ``4d conformal theories and strings on orbifolds,''
  Phys.\ Rev.\ Lett.\  {\bf 80}, 4855 (1998)
  [arXiv:hep-th/9802183].
  
\bibitem{morrison}
  D.~R.~Morrison and M.~R.~Plesser,
  ``Non-spherical horizons. I,''
  Adv.\ Theor.\ Math.\ Phys.\  {\bf 3}, 1 (1999)
  [arXiv:hep-th/9810201].
         
\bibitem{Benvenuti} 
  S.~Benvenuti, S.~Franco, A.~Hanany, D.~Martelli and J.~Sparks,
  ``An Infinite family of superconformal quiver gauge theories with Sasaki-Einstein duals,''
  JHEP {\bf 0506}, 064 (2005)
  [hep-th/0411264].
  
\bibitem{benvenuti} 
  S.~Benvenuti, M.~Mahato, L.~A.~Pando Zayas and Y.~Tachikawa,
  ``The Gauge/gravity theory of blown up four cycles,''
  hep-th/0512061.
  
\bibitem{Aharony} 
  O.~Aharony, O.~Bergman, D.~L.~Jafferis and J.~Maldacena,
  ``N=6 superconformal Chern-Simons-matter theories, M2-branes and their gravity duals,''
  JHEP {\bf 0810}, 091 (2008)
  [arXiv:0806.1218 [hep-th]].
         
\bibitem{Ross} 
  S.~F.~Ross,
  ``Non-supersymmetric asymptotically AdS$_5\times S^5$ smooth geometries,''
  JHEP {\bf 0601}, 130 (2006)
  [hep-th/0511090].
        
\bibitem{Buchel} 
  A.~Buchel and J.~T.~Liu,
  ``Gauged supergravity from type IIB string theory on $Y^{p,q}$ manifolds,''
  Nucl.\ Phys.\ B {\bf 771}, 93 (2007)
  [hep-th/0608002].
                 
\bibitem{Murugan} 
  I.~R.~Klebanov and A.~Murugan,
  ``Gauge/Gravity Duality and Warped Resolved Conifold,''
  JHEP {\bf 0703}, 042 (2007)
  [hep-th/0701064].

\bibitem{Cvetic-JFVP} 
  M.~Cvetic and J.~F.~Vazquez-Poritz,
  ``Warped resolved $L^{a,b,c}$ cones,''
  Phys.\ Rev.\ D {\bf 77}, 126003 (2008)
  [arXiv:0705.3847 [hep-th]].

\bibitem{calabi}
  E. Calabi, ``M\'etriques kahl\'eriennes et fibr\'es holomorphes,'' Ann. Sci. Ecole Normale Super. {\bf 12}, 269 (1979).

\bibitem{page}
  D.~N.~Page, C.~N.~Pope,
  ``Inhomogeneous Einstein Metrics On Complex Line Bundles,''
  Class.\ Quant.\ Grav.\  {\bf 4}, 213-225 (1987).
        
\bibitem{cornish}
  N.~J.~Cornish, D.~N.~Spergel and G.~D.~Starkman,
  ``Circles in the Sky: Finding Topology with the Microwave Background
  Radiation,''
  Class.\ Quant.\ Grav.\  {\bf 15}, 2657 (1998)
  [arXiv:astro-ph/9801212].
    
\bibitem{becker}
  K.~Becker,
  ``A Note on compactifications on spin(7) - holonomy manifolds,''
  JHEP {\bf 0105}, 003 (2001).
  [hep-th/0011114].

\bibitem{untwisted} 
  M.~J.~Duff, H.~Lu and C.~N.~Pope,
  ``AdS$_5\times S^5$ untwisted,''
  Nucl.\ Phys.\ B {\bf 532}, 181 (1998)
  [hep-th/9803061].

\bibitem{pando}
  L.~A.~Pando Zayas, A.~A.~Tseytlin,
  ``3-branes on spaces with $R\times S^2\times S^3$ topology,''
  Phys.\ Rev.\  {\bf D63}, 086006 (2001).
  [hep-th/0101043].

\bibitem{Candelas} 
  P.~Candelas and X.~C.~de la Ossa,
  ``Comments on Conifolds,''
  Nucl.\ Phys.\ B {\bf 342}, 246 (1990).

\bm{berard}
L. Berard-Bergery,
``Quelques examples de varietes riemanniennes completes non-compactes a courbure de Ricci positive," C.R. Acad. Sci. Ser. {\bf I302}, 159 (1986).

\bibitem{klebanov-witten}
  I.~R.~Klebanov and E.~Witten,
  ``Superconformal field theory on threebranes at a Calabi-Yau  singularity,''
  Nucl.\ Phys.\  B {\bf 536}, 199 (1998)
  [arXiv:hep-th/9807080].

\bibitem{Gauntlett1} 
  J.~P.~Gauntlett, D.~Martelli, J.~Sparks and D.~Waldram,
  ``Supersymmetric AdS$_5$ solutions of M theory,''
  Class.\ Quant.\ Grav.\  {\bf 21}, 4335 (2004)
  [hep-th/0402153].

\bibitem{Gauntlett2} 
  J.~P.~Gauntlett, D.~Martelli, J.~Sparks and D.~Waldram,
  ``Sasaki-Einstein metrics on $S^2\times S^3$,''
  Adv.\ Theor.\ Math.\ Phys.\  {\bf 8}, 711 (2004)
  [hep-th/0403002].

\bibitem{Pal} 
  S.~S.~Pal,
  ``A New Ricci flat geometry,''
  Phys.\ Lett.\ B {\bf 614}, 201 (2005)
  [hep-th/0501012].

\bibitem{Sfetsos} 
  K.~Sfetsos and D.~Zoakos,
  ``Supersymmetric solutions based on $Y^{p,q}$ and $L^{p,q,r}$,''
  Phys.\ Lett.\ B {\bf 625}, 135 (2005)
  [hep-th/0507169].

\bibitem{Cvetic1} 
  M.~Cveti\v{c}, H.~L\"u, D.~N.~Page and C.~N.~Pope,
  ``New Einstein-Sasaki spaces in five and higher dimensions,''
  Phys.\ Rev.\ Lett.\  {\bf 95}, 071101 (2005)
  [hep-th/0504225].

\bibitem{orbifold1} 
  L.~J.~Dixon, J.~A.~Harvey, C.~Vafa and E.~Witten,
  ``Strings on Orbifolds,''
  Nucl.\ Phys.\ B {\bf 261}, 678 (1985).

\bibitem{orbifold2} 
  L.~J.~Dixon, J.~A.~Harvey, C.~Vafa and E.~Witten,
  ``Strings on Orbifolds. 2.,''
  Nucl.\ Phys.\ B {\bf 274}, 285 (1986).

\bibitem{Oota} 
  T.~Oota and Y.~Yasui,
  ``Explicit toric metric on resolved Calabi-Yau cone,''
  Phys.\ Lett.\ B {\bf 639}, 54 (2006)
  [hep-th/0605129].

\bibitem{Horowitz} 
  G.~T.~Horowitz and R.~C.~Myers,
  ``The AdS / CFT correspondence and a new positive energy conjecture for general relativity,''
  Phys.\ Rev.\ D {\bf 59}, 026005 (1998)
  [hep-th/9808079].
  
\bibitem{Duff} 
  M.~Cveti\v{c}, M.~J.~Duff, P.~Hoxha, J.~T.~Liu, H.~L\"u, J.~X.~Lu, R.~Martinez-Acosta and C.~N.~Pope {\it et al.},
  ``Embedding AdS black holes in ten and eleven dimensions,''
  Nucl.\ Phys.\ B {\bf 558}, 96 (1999)
  [hep-th/9903214].
  
\bibitem{Hoxha} 
  P.~Hoxha, R.~R.~Martinez-Acosta and C.~N.~Pope,
  ``Kaluza-Klein consistency, Killing vectors, and Kahler spaces,''
  Class.\ Quant.\ Grav.\  {\bf 17}, 4207 (2000)
  [hep-th/0005172].

\bibitem{Hawking} 
  S.~W.~Hawking,
  ``Gravitational Instantons,''
  Phys.\ Lett.\ A {\bf 60}, 81 (1977).

\bibitem{Awad} 
  A.~Awad and A.~Chamblin,
  ``A Bestiary of higher dimensional Taub-NUT AdS space-times,''
  Class.\ Quant.\ Grav.\  {\bf 19}, 2051 (2002)
  [hep-th/0012240].

\bibitem{Bena} 
  I.~Bena, S.~Giusto, C.~Ruef and N.~P.~Warner,
  ``A (Running) Bolt for New Reasons,''
  JHEP {\bf 0911}, 089 (2009)
  [arXiv:0909.2559 [hep-th]].

\bibitem{Chamblin} 
  A.~Chamblin, R.~Emparan, C.~V.~Johnson and R.~C.~Myers,
  ``Large N phases, gravitational instantons and the nuts and bolts of AdS holography,''
  Phys.\ Rev.\ D {\bf 59}, 064010 (1999)
  [hep-th/9808177].

\bibitem{page-pope3} 
  H.~L\"u, D.~N.~Page and C.~N.~Pope,
  ``New inhomogeneous Einstein metrics on sphere bundles over Einstein-Kahler manifolds,''
  Phys.\ Lett.\ B {\bf 593}, 218 (2004)
  [hep-th/0403079].

\bibitem{Mann} 
  R.~B.~Mann and C.~Stelea,
  ``New multiply nutty spacetimes,''
  Phys.\ Lett.\ B {\bf 634}, 448 (2006)
  [hep-th/0508203].

\bibitem{Singh} 
  H.~Singh,
  ``M2-branes on a resolved $C_4/Z_4$,''
  JHEP {\bf 0809}, 071 (2008)
  [arXiv:0807.5016 [hep-th]].

\bibitem{Krishnan} 
  C.~Krishnan, C.~Maccaferri and H.~Singh,
  ``M2-brane Flows and the Chern-Simons Level,''
  JHEP {\bf 0905}, 114 (2009)
  [arXiv:0902.0290 [hep-th]].
 
\bibitem{DAuria} 
  R.~D'Auria, P.~Fre and P.~van Nieuwenhuizen,
  ``N=2 Matter Coupled Supergravity From Compactification On A Coset G/h Possessing An Additional Killing Vector,''
  Phys.\ Lett.\ B {\bf 136}, 347 (1984).

\bibitem{Fabbri} 
  D.~Fabbri, P.~Fre', L.~Gualtieri, C.~Reina, A.~Tomasiello, A.~Zaffaroni and A.~Zampa,
  ``3-D superconformal theories from Sasakian seven manifolds: New nontrivial evidences for AdS$_4$ / CFT$_3$,''
  Nucl.\ Phys.\ B {\bf 577}, 547 (2000)
  [hep-th/9907219].
  
\bibitem{Franco} 
  S.~Franco, I.~R.~Klebanov and D.~Rodriguez-Gomez,
  ``M2-branes on Orbifolds of the Cone over $Q^{1,1,1}$,''
  JHEP {\bf 0908}, 033 (2009)
  [arXiv:0903.3231 [hep-th]].

\bibitem{Amariti} 
  A.~Amariti, D.~Forcella, L.~Girardello and A.~Mariotti,
  ``3D Seiberg-like Dualities and M2-Branes,''
  JHEP {\bf 1005}, 025 (2010)
  [arXiv:0903.3222 [hep-th]].
  
\bibitem{Davey} 
  J.~Davey, A.~Hanany, N.~Mekareeya and G.~Torri,
  ``Phases of M2-brane Theories,''
  JHEP {\bf 0906}, 025 (2009)
  [arXiv:0903.3234 [hep-th]].

  \bibitem{stenzel}
  M.B. Stenzel,
  ``Ricci-flat metrics on the complexification of a compact rank one symmetric space,''
  Manuscripta Mathematica {\bf 80}, 151 (1993).
  [hep-th/0011114].

\bibitem{bergman}
  A.~Bergman, C.~P.~Herzog,
  ``The Volume of some nonspherical horizons and the AdS/CFT correspondence,''
  JHEP {\bf 0201}, 030 (2002).
  [hep-th/0108020].

\bibitem{Martelli2} 
  D.~Martelli and J.~Sparks,
  ``AdS$_4$ / CFT$_3$ duals from M2-branes at hypersurface singularities and their deformations,''
  JHEP {\bf 0912}, 017 (2009)
  [arXiv:0909.2036 [hep-th]].

\bibitem{Witten} 
  E.~Witten,
  ``Search for a Realistic Kaluza-Klein Theory,''
  Nucl.\ Phys.\ B {\bf 186}, 412 (1981).

\bibitem{7D1} 
  J.~P.~Gauntlett, D.~Martelli, J.~F.~Sparks and D.~Waldram,
  ``A New infinite class of Sasaki-Einstein manifolds,''
  Adv.\ Theor.\ Math.\ Phys.\  {\bf 8}, 987 (2006)
  [hep-th/0403038].
  
\bibitem{7D2} 
  W.~Chen, H.~L\"u, C.~N.~Pope and J.~F.~V\'azquez-Poritz,
  ``A Note on Einstein Sasaki metrics in $D\ge 7$,''
  Class.\ Quant.\ Grav.\  {\bf 22}, 3421 (2005)
  [hep-th/0411218].

\bibitem{7D3} 
  H.~L\"u, C.~N.~Pope and J.~F.~V\'azquez-Poritz,
  ``A New construction of Einstein-Sasaki metrics in $D\ge 7$,''
  Phys.\ Rev.\ D {\bf 75}, 026005 (2007)
  [hep-th/0512306].

\bibitem{Martelli} 
  D.~Martelli and J.~Sparks,
  ``Moduli spaces of Chern-Simons quiver gauge theories and AdS$_4$/CFT$_3$,''
  Phys.\ Rev.\ D {\bf 78}, 126005 (2008)
  [arXiv:0808.0912 [hep-th]].

\bibitem{Hanany} 
  A.~Hanany and A.~Zaffaroni,
  ``Tilings, Chern-Simons Theories and M2-Branes,''
  JHEP {\bf 0810}, 111 (2008)
  [arXiv:0808.1244 [hep-th]].

\bibitem{Ueda} 
  K.~Ueda and M.~Yamazaki,
  ``Toric Calabi-Yau four-folds dual to Chern-Simons-matter theories,''
  JHEP {\bf 0812}, 045 (2008)
  [arXiv:0808.3768 [hep-th]].

\bibitem{Eguchi} 
  T.~Eguchi and A.~J.~Hanson,
  ``Asymptotically Flat Selfdual Solutions to Euclidean Gravity,''
  Phys.\ Lett.\ B {\bf 74}, 249 (1978).

\bibitem{Taub} 
  A.~H.~Taub,
  ``Empty space-times admitting a three parameter group of motions,''
  Annals Math.\  {\bf 53}, 472 (1951).

\end{thebibliography}
\end{document}